\documentclass[a4paper,11pt]{article}

\usepackage[utf8]{inputenc}
\usepackage{cancel}
\usepackage{tikz}
\usepackage{ulem}
\usepackage{amsfonts}
\usepackage{amssymb}
\usepackage{graphicx}
\usepackage{amsmath}
\usepackage{enumerate}
\usepackage{mathtools}
\usepackage{subfig}
\usepackage{color}
\usepackage{tikz}
\usepackage{float}
\usepackage{here}
\usepackage{cite}
\usepackage{mathrsfs}
\usepackage{float,epsfig}
\usepackage{dcolumn}
\usepackage{graphicx}
\usepackage{bm}
\usepackage{booktabs}
\usepackage{multirow}
\usepackage{changepage}
\usepackage{amsmath,amssymb,amsthm}
\usepackage[colorlinks=true,linkcolor=blue,citecolor=red]{hyperref}
\usepackage{booktabs} 
\usepackage{siunitx}  
\tikzset{every picture/.style={line width=0.75pt}} 
\usetikzlibrary{arrows.meta, positioning}
\usepackage{multirow}
\usepackage[toc,page]{appendix}
\usetikzlibrary{arrows.meta}
\usetikzlibrary{bending}
\usetikzlibrary{calc}
\newcommand{\be}{\begin{equation}}
\newcommand{\ee}{\end{equation}}
\newcommand{\bea}{\setlength\arraycolsep{2pt} \begin{eqnarray}}
\newcommand{\eea}{\end{eqnarray}}

\def\0{{\sst{(0)}}}
\def\1{{\sst{(1)}}}
\def\2{{\sst{(2)}}}
\def\3{{\sst{(3)}}}
\def\4{{\sst{(4)}}}
\def\5{{\sst{(5)}}}
\def\6{{\sst{(6)}}}
\def\7{{\sst{(7)}}}
\def\8{{\sst{(8)}}}
\def\sst#1{{\scriptscriptstyle #1}}

\makeatletter \@addtoreset{equation}{section}

\definecolor{lime}{HTML}{A6CE39}

\begin{document}
\title{{\normalsize \textbf{\Large  On   Thermodynamic Universalities  and Optics  of   $B$-deformed  Reissner--Nordstr\"om--AdS  Black Holes}}}
\author{ {\small   Adil Belhaj and  Maryem  Jemri\thanks{Corresponding author: maryem.jemri@um5r.ac.ma} \footnote{
Authors are listed  in alphabetical order. 
} \hspace*{-8pt}} \\
{\small  ESMaR, Faculty of Science, Mohammed V University in Rabat, Rabat, Morocco  }}
\maketitle

\begin{abstract}
Using machine learning techniques, we    study the  Reissner--Nordstr\"om--AdS  black holes  in the framework of   extended space-time  derivatives inspired  by  non-commutative  geometry  in string theory   where   the  deformation parameters are related to  the inverse of the NS–NS $B$-field.  We refer to these solutions   as  $B$-deformed Reissner--Nordstr\"om--AdS  black holes.  Precisely, we examine    the  thermodynamic and the optical  behaviors  of such   stringy deformed  AdS black holes.   Exploiting  advanced numerical computations,  we first  examine the  corresponding  thermodynamic Van der Waals behaviors by  calculating the  universal ratios  describing the $P$--$V$ criticality and the  Joule–Thomson expansion features. 
Applying  such  numerical results, we   construct  reliable training data for a fully connected neural network being   developed  to determine whether such stringy    black hole configurations exhibit  thermodynamic Van der Waals aspects.   Concretely, we reveal  that the fully connected and trained neural network accurately detects Van der Waals behaviors.    Supported by  the observational data reported by the Event  Horizon Telescope international  collaborations, we    investigate the optical shadows  of such  $B$-deformed Reissner--Nordstr\"om--AdS  black holes using machine learning methods.   In order to provide corroborated models, we constraint the involved    parameters including  the stringy one $B$   by making use of   M87* and Sgr A*   black hole empirical data.

\end{abstract}

\newpage

\section{Introduction}
The study of black holes has revealed the existence of a bridge between  gravity theory  and statistical mechanics including   condensed matter physics where  the concept of analogous black holes has been implemented\cite{r4,r5,r6,DFT1,DFT2}.  In fact,  it has been suggested that the low excitation of electrons  could behave like particles in high energy physics.    Alternatively, the black hole  activities have  been corroborated by the results of the Event Horizon Telescope (EHT), which succeeded in capturing the first-ever image of the shadow of a black hole in M87*, and then in Sgr A* \cite{r7,r8}. Thanks to such  recent observational discoveries,  the black hole parameters can be checked using  the optical data via shadow configurations. In this way, the geometric parameters, including the size and  the shape of the  involved  shadows,  can be  exploited  to estimate the relevant quantities   of the black holes   including the ones appearing in  the  extended  models  of gravity supported by non-trivial physical theories  \cite{r11,r12}.  Correlating theoretical predictions of the  black hole shadows with EHT  data contributes  to  provide certain   constraining scenarios of  the involved  moduli  space.

Alternative  investigations  have focused on the  thermodynamic behaviors  establishing    relevant similarities to physical systems undergoing universal phase transitions \cite{i,ii}. By interpreting the cosmological constant as a thermodynamic pressure, it has been shown that  the   Anti de Sitter  (AdS)   black holes exhibit Van der Waals-type  statistical behaviors  with $P$–$V$ criticality  aspects  \cite {r9,ad4}.    These universal features connect the microscopic structure of space-times to  the macroscopic thermodynamic quantities. This  interplay  has been used to develop a deeper understanding of gravitational systems leading  to the emergence of  certain  universal constants in AdS  black hole physics\cite{r99}.

Recently, machine learning has become a powerful tool in the field of the  black hole physics. Among  others,  it  enables the extraction of physical parameters, the classification of black holes, and the improvement of image reconstructions from empirical  data. Neural networks (NNs), for instance,  can be used to determine  the accretion parameters from horizon-scale images \cite{r101}.   In addition, they  clarify  images of black holes obtained by the EHT international  collaborations \cite{r102}. They have also been exploited  to  classify black holes in the mass-spin diagram \cite{r103}. These studies have demonstrated potential applications   of machine learning and analyzing complex measurement sets.

More recently,  the Schwarzschild--AdS black holes in non-commutative (NC) space-times have attracted remarkable  interest, particularly in the study of their thermodynamic properties and phase transitions \cite{n1}. Among the various realizations of NC space-times, Lie-algebraic deformations  explored in the framework of doubly special relativity  generate a natural description  for  considering  NC  corrections to black hole physics \cite{n2}. The $\kappa$-deformed scenario, for instance, marked by a Lie-algebraic deformation of the space-time coordinates, has been utilized  to investigate the thermodynamics of  Schwarzschild--AdS black holes. It has been shown that  this  $\kappa$-deformation modifies the thermodynamic quantities of the ordinary Schwarzschild solution  providing  non-trivial corrections to the  state equation  and the  phase structure.  Precisely,  such  $\kappa$-deformed corrections have been revealed  to exhibit thermodynamic behaviors analogous to that of Van der Waals fluids, showing  characteristic critical aspects and phase transition behaviors \cite{n3,n4}.

Using machine learning techniques, we  contribute to these activities by  investigating   the  thermodynamic and the optical  behaviors   of the  Reissner--Nordstr\"om--AdS (RN-AdS)  black holes   in the framework of  extended space-time  derivatives inspired  by NC geometry  in string theory   where  the  NC parameters are linked the inverse of the   antisymmetric tensor field $B$ of the NS-NS sector of type II superstrings.  We refer to these  solutions   as  $B$-deformed RN–AdS black holes. Utilizing   advanced numerical methods,  we first  assess  the  thermodynamic Van der Waals behaviors by  calculating the  universal ratios  describing   the $P$--$V$ criticality and the  Joule–Thomson expansion aspects. 
Employing  these numerical results, we   build   training data from such numerical results  for a fully connected neural network (FCNN) designed to determine whether such   black hole configurations   exhibit thermodynamic Van der Waals behaviors.  Precisely, we show that the fully connected and trained neural network (NN) accurately detects Van der Waals behaviors.  Encouraged  by empirical findings   reported by the EHT international  collaborations, we    study the optical shadows of such stringy deformed solutions   using machine learning methods.  In order to supply corroborated models, we constraint the involved  parameters including  the stringy one $B$ with the help of    M87* and Sgr A*    black hole  observational data. 

This work is organized as follows.  In section 2, we present a  concise  discussion on   $B$-deformed charged AdS black holes motivated by NC  geometry  in string theory. Section 3  is devoted to  the investigation of  certain   Van der Waals  universals and constraining parameters via thermodynamics    using  machine learning methods.     In section 4, we use  such numerical computation scenarios to classify whether  such black holes   match with EHT collaboration findings.  The last section provides   concluding remarks.

\section{ B-deformed  RN-AdS  black holes}
As mentioned, RN-AdS black holes  have received a particular  interest in connection with thermodynamics and  other issues including  string theory and NC geometry activities\cite{nn1,nn2}.  It is recalled that  NC  geometry  arises  from D-brane behaviors being  coupled to the closed string  spectrum including the metric $g_{\mu\nu}$ and the antisymmetric tensor  $B_{\mu\nu}$.  In the  Seiberg-Witten limit,   the correlation functions of the one dimensional boundary string fields usually verify  certain 
non-trivial commutation relations  producing  a NC space-time where  the star product of the 
Moyal bracket  has been used in real functions  spaces associated with open string ends living in D-brane objects \cite{nc1,nc2,SM1}.  In thermodynamics,  the cosmological constant   $\Lambda$ has been considered as a thermodynamic  pressure  $P$ via the relation
\begin{equation}
P=- \frac{\Lambda}{8\pi}.
\end{equation}
This link has been  largely exploited to approach a class of static, spherically symmetric space-time given by the following  line element    form 
\begin{equation}
ds^2=g_{\mu\nu}dx^\mu dx^\nu=-f(r)\,dt^2+\frac{1}{f(r)}\,dr^2+r^2\left(d\theta^2+\sin^2\theta\, d\phi^2\right).
\label{eq:metric}
\end{equation}
 In the ordinary solutions, the metric function  takes the form  
\begin{equation}
f(r)=1-\frac{2M}{r}+\frac{Q^2}{r^2}-\frac{\Lambda}{3}r^2.
\label{eqo}
\end{equation}
In this radial function,  $M$, $Q$, and $\Lambda$ denote  the mass,  the  charge, and  the cosmological constant, respectively.  Generalized solutions  have been extensively  studied in order to  investigate     physical behaviors  of certain  black hole models   which could be corroborated via falsification scenarios. A close examination shows that  these extended models involve extra parameters usually called external ones. They are either inspired or supported by non-trivial gravity theories with exotic matter and energies. In  certain appropriate discussions, the metric function of such modified gravity theories  could  be written as   follows
\begin{equation}
f_d(r)=1-\frac{2M_d(r)}{r}+\frac{Q^2_d(r)}{r^2}-\frac{\Lambda_d(r)}{3}r^2.
\label{eqd}
\end{equation}
Up to some parameter  considerations, one could write
\begin{eqnarray}
M_d(r)&=& M(1-m(r))\\
Q^2_d(r)&=&Q^2(1-q(r)) \\
\Lambda_d(r)&=& \Lambda (1-\lambda(r))
\end{eqnarray}
where $m(r)$,  $q(r)$, and  $\lambda(r)$ are dimensionless  radial functions  carrying  information on   the involved external parameters. The associated explicit forms can be obtained  by solving the   Einstein  equations  of motion  derived  from the action of the theory in question.  Several scenarios have  been  provided   generating non-trivial forms of such radial functions with external parameters supported by the associated spectrum fields. Here, however, we reconsider the study of  stingy solutions derived by implementing a new type of  space-ime derivatives  inspired by NC  geometry in string theory where the deformation  parameters are linked to the inverse of the B-field of the  NS-NS sector appearing in type II superstrings\cite{nc1}.   Concretely,  they are given by 
\begin{eqnarray}\label{}
	D_{x_{\mu}}= \frac{\partial}{\partial x_{\mu}}+ B_{\mu\nu}x^\nu
	\end{eqnarray}
where  $B_{\mu\nu}$ is  a  real  tensor  of  rank 2  shearing  similarities with  nonzero  antisymmetric tensor called $B$-field  associated with  fundamental string objects in string theory  \cite{ref1}. Considering only  space backgrounds, we take the following form  
\begin{eqnarray}\label{}
D_{t}=\frac{\partial}{\partial t},\qquad 
	D_{x_{i}}= \frac{\partial}{\partial x_{i}}+ B_{ij}x^j \quad\quad i,j=1,2,3
	\end{eqnarray}
where now   $B_{ij}$ is  a  real  tensor  of  rank 2	 in three dimensions.  For simplicity reasons,  we  deal with the following   economical  form  \begin{equation}
B_{ij}= B \epsilon_{ij},
\end{equation}
where  $\epsilon^{ij}$ is the usual antisymmetric tensor of order 2  and 
 $B$  is  a   real  free  parameter of dimension $ [L^{-2}] $ being relevant   in the present investigation, we refer to as stringy parameter.    According to \cite{ref1},   it has been shown that the above  deformation modifies the standard gravitational background through an additional contribution to the ordinary metric function.  The latter has been found to be 
\begin{equation}
f(r)=1-\frac{2M}{r}+\frac{Q^2}{r^2}-\frac{\Lambda}{3}r^2
+\frac{2Br^2}{3}\log\!\left(r/r_0\right),
\label{eq:fg}
\end{equation}
where one has used 
\begin{eqnarray}
m(r)=0\qquad 
q(r)=0 \qquad
\lambda(r)=  \frac{2B}{\Lambda}\log\!\left(r/r_0\right).
\end{eqnarray}
Here,  $r_0$ has a length dimension introduced to recover the dimensionless logarithmic function, which we can consider equal to one without loss of generality. The resulting geometry can be interpreted as a deformation of the RN--AdS space-time, where deviations from the standard solution arise naturally from the tensorial deformation encoded in the antisymmetric tensor  $B_{ij}$.   We refer to this  as  $B$-deformed RN–AdS black holes.

\begin{figure}[h!]
\centering
\includegraphics[width=0.5\textwidth]{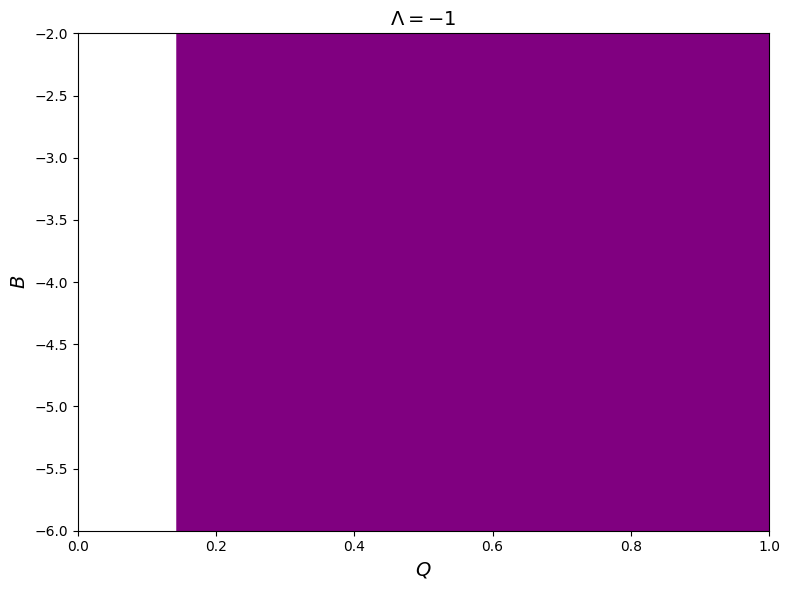} 
 \caption{ Regions in the $(B, Q)$ plane where the metric admits at least one real event horizon
radius with $M=1$.
} \label{22}
\end{figure}

Roughly, the horizons are identified by vanishing the inverse of the radial component of the metric function. To probe their existence,
we employ a numerical method in which the parameters M,  and $\Lambda$ remain constant, while
the parameter $B$ varies from $-4$ to $-2$ in increments of 0.1.   For each pair $(B,Q)$, the horizon equation is
solved numerically in order to  determine whether there exists at least one real root corresponding to a physically meaningful horizon in black hole physics. This systematic procedure
provides a way to identify the regions of the reduced black hole moduli space in which these
solutions are acceptable. Fig. (\ref{22}) illustrates such behaviors by displaying the regions in the
$(B,Q)$ plane where at least one real horizon exists. 

Increasing the parameter $Q$ reduces the regions of the moduli 
space which supports the existence of physical horizons. Based on these results, the forthcoming  discussion  will be conducted within the parameter subsets which correspond to
configurations admitting at least one real horizon in the associated black hole physics.

\section{Thermodynamic universalities  of  $B$-deformed  RN-AdS black holes using  machine learning  computations }
In this section, we investigate the universal behaviors of such  $B$-deformed  RN-AdS black holes using machine learning methods and  accelerated  numerical computations. Precisely, we examine their criticality and universal properties. First, we derive constraints on the deformed black hole parameters with the help of  Van der Waals  behaviors using accelerated  numerical  methods. After that,   we  analyze  the Van der Waals  properties  of such stringy  black holes by using  machine learning techniques with  accelerated  numerical  methods.  It is recalled that accelerated  numerical methods  enable  efficient parallel computing by taking advantage of NVIDIA GPU architecture \cite{rr}.  It distributes tasks across multiple streaming multiprocessors (SMs), providing high-performance parallel processing and advanced tools to optimize computations and simulation scenarios.
\subsection{Constraining  parameters from Van der Waals  behaviors}
In the first part, we investigate the $P$-$V$ criticality and the Joule-Thomson expansion of  such  stringy  deformed black hole solutions. 
To  do so,  certain  thermodynamic quantities are needed to be  assessed.  Indeed, we start by discussing the $P$-$V$ criticality behaviors of such  deformed charged-like  AdS black hole solutions. The associated  thermodynamic pressure   has been  found to be 
\begin{equation}
P = \frac{6B \ln(r_h)\, r_h^4 + 2B r_h^4 + 12 T r_h^3 - 3 Q^2 + 3 r_h^2}{24 r_h^4}
\end{equation}
where $T$ is the Hawking temperature.  Introducing the specific volume $v = 2 r_h$, the equation of state  can be written as  follows
\begin{equation}
P = B\frac{\log(2) - \log(v)}{4} + \frac{B}{12} + \frac{T}{v} + \frac{2 Q^2}{v^4} - \frac{1}{2 v^2}.
\end{equation}
The critical point $(P_c, T_c, v_c)$  can  be  determined by solving  the inflection point conditions
\begin{equation}
\frac{\partial P}{\partial v} = 0, 
\qquad 
\frac{\partial^2 P}{\partial v^2} = 0,
\end{equation}
as discussed in~\cite{ref1}.  Usually, it  generates a universal  ratio  which can be expressed as follows 
\begin{equation}
\chi=\dfrac{P_{c}v_{c}}{T_{c}}. 
\end{equation}
After computations,  the critical pressure, the  temperature, and the specific volume  have been  demonstrated to be 
\begin{eqnarray}
P_c&=&\frac{3  B\left(12 B \,Q^{2}+\xi \right)  \left(\ln \! \left(2\right)-2 \ln \! \left(-\frac{\sqrt{-B \xi }}{B}\right)\right)-114 B^{2} Q^{2}-11 B \xi }{12 \xi ^{2} \pi} \nonumber\\ 
T_{c}&=&-\frac{  \sqrt{2} B\left(8 B \,Q^{2}+\xi \right) }{\xi  \sqrt{-B \xi }\, \pi}\\
v_{c}&=&-\frac{\sqrt{-2 B \xi }}{B}  \nonumber
\end{eqnarray}
where  one has used $\xi=\xi(B,Q)=1+\sqrt{1+24 B \,Q^{2}}$.  It has been observed that certain constraints must be imposed on the stringy  parameter $B$
 in order to obtain real and physically meaningful critical quantities. Real  quantities  require the constraint $
-\dfrac{1}{24Q^2}< B < 0$ 
 providing a  range  of negative values for the stringy parameter $B$.  Indeed, the  critical  values provide the following ratio
\begin{equation}
\chi=\frac{
(36 B Q^2 + 3\xi)\left(-\ln(2) + 2\ln\left(-\frac{\sqrt{-B \xi}}{B}\right)\right) + 114 B Q^2 + 11\xi 
}{
96 B Q^2 + 12\xi
}.
\end{equation}
It has been observed  that, unlike the  ordinary charged AdS black holes,  $\chi$ is not a universal constant. To further explore the critical   behaviors  of the proposed black hole solutions, we  examine  the Joule--Thomson expansion behaviors. 
For fixed charge values, the Joule--Thomson coefficient  is expressed as 
\begin{equation}
\mu=\left( \dfrac{\partial T}{\partial P} \right)_{M}
=\dfrac{1}{C_{P}} \left[ T \left( \dfrac{\partial V}{\partial T} \right)_{P}-V  \right].
\end{equation}
Imposing the inversion condition $\mu(T_i)=0$ together with the thermodynamic volume leads to the inversion temperature
\begin{equation}
T_i=\frac{6 r_i^{4} B \ln \! \left(r_i^{3}\right)+8 \left(3 \pi  P_i +B \right) r_i^{4}-3 r_i^{2}+9 Q^{2}}{36 r_i^{3} \pi}.
\end{equation}
Using the equation of state
\begin{equation}
T=\frac{6 B \ln \! \left(r \right) r^{4}+\left(24 \pi  P +2 B \right) r^{4}+3 r^{2}-3 Q^{2}}{12 r^{3} \pi},
\end{equation}
and setting $T=T_i$, one  can obtain a constraint relating the inversion pressure $P_i$ and the horizon radius $r_i$.  Indeed,  the minimal inversion temperature is shown to be 
\begin{equation}
T_{i}^{min}=\frac{\sqrt{3}\, \left(\left(B \,Q^{2}-2\kappa \right) \ln \! \left(\frac{\kappa}{B}\right)-2 \ln \! \left(3\right)-\frac{5}{3}+B \ln \! \left(3\right) Q^{2}+\frac{7 B \,Q^{2}}{9}\right)}{-4 \pi \kappa \sqrt{\frac{\kappa}{B}}\, },
\end{equation}
where we have used $
\kappa=1-\sqrt{1-BQ^2}$.  This  helps to generate  the  following ratio
\begin{equation}
\zeta= \dfrac{T_{i}^{min}}{T_{c}}. 
\end{equation}
The calculations give 
 \begin{equation} 
 \zeta =\frac{\xi\sqrt{-B \xi}\, \left(\left(B \,Q^{2}-2\kappa\right) \ln \! \left(\frac{\kappa}{B}\right) \ln \! \left(3\right)-\frac{5}{3}+B \ln \! \left(3\right) Q^{2}+\frac{7 B \,Q^{2}}{9}\right) \sqrt{6}}{8 B\sqrt{\frac{\kappa}{B}}\, \kappa \left(8 B \,Q^{2}+\xi\right) }.
  \end{equation}
The resulting expression shows that $\zeta$ is not constant, in contrast with the behavior appearing  in  the ordinary  RN-AdS black holes. In this way, the thermodynamic behaviors depend strongly on the stringy deformation parameter $B$ and the electric charge $Q$.  For  certain parameter values, we could obtain the reference values $\chi=\frac{3}{8}$ and $\zeta=\frac{3}{4}$, recovering the standard Van der Waals behaviors   and the Van der Waals--like Joule--Thomson expansion of charged AdS solutions, respectively

\begin{figure}[h!]
\centering
\includegraphics[width=0.6\textwidth]{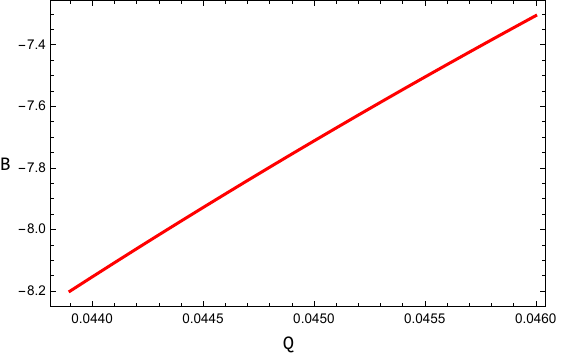}
\caption{$B$ as a function of $Q$ producing Van der Waals behaviors.}
\label{4}
\end{figure}

In order to support the  Van der Waals-like behaviors  for such stringy  charged AdS  black holes, it is necessary to determine the couple  parameters $(B,Q)$ satisfying  the  universal thermodynamic ratios for both the  $P$--$V$ criticality and the  Joule–Thomson expansion features. 
In fact, extensive numerical computations are  needed  to explore such parameter space constraints  by  identifying the  physical relevant regions  in the black hole moduli space.  To  classify and predict the domains that exhibit    such  universality requirements, we can  exploit  machine learning techniques methods. In the study of black holes,  parallel programing  significantly reduces  the computation time and improves the numerical stability. It has been used to simulate black hole shadows, linking theoretical models to observations from the EHT findings \cite{cuda,C31}. Combined with machine learning,  it  enables the analysis of large data sets from complex simulations, resulting in faster and more accurate predictions of black hole properties. In this context, we move  now to  investigate the critical limits of Van der Waals features for the proposed stringy  black holes. In criticality studies, such combined behaviors can be achieved by setting the following  ratio values
\begin{equation}
\chi = \frac{3}{8},\qquad
\zeta = \frac{3}{4}.
\end{equation}
A close examination shows that these conditions impose rigid limits on the allowed ranges of the parameters, called critical values. To find the matching couples $(B, Q)$ in the reduced moduli space, extra requirements on the stringy parameter $B$ should be imposed. Precisely, we consider  the conditions $
B < 0$ and $24BQ^2+1 > 0$. To conduct this analysis, we have developed a dedicated accelerated  numerical code. First, we generate a set of candidate pairs $(B,Q)$ within the physically allowed ranges of the parameters, restricting $Q$ to range from $0$ to $0.082$ with a  discretization step $\Delta Q = 0.001$. Next, we evaluate each pair to determine whether it simultaneously satisfies the $P$-$V$ criticality condition $\chi = \frac{3}{8}$ and the Joule--Thomson expansion condition $\zeta = \frac{3}{4}$.  The pairs fulfilling both constraints are retained and represented graphically as a function of $B$ versus $Q$  as illustrated in Fig.(\ref{4}). In fact,  each point  of such a curve corresponds to a location in the reduced moduli space exhibiting the Van der Waals behaviors. 

\subsection{ Machine learning  study  of Van der Waals behaviors}

The parallel programming  exploration has identified regions of  the parameters exhibiting  the Van der Waals behaviors. To analyze these results in more detail, we apply machine learning techniques to the data generated by such numerical  methods. This approach enables automatic classifications and predictions of the  Van der Waals behaviors in the parameter space of the  black holes. This methodology has been inspired by recent applications of machine learning in black hole physics and string theory, where NNs have been successfully used to classify black holes in the mass-spin diagram and to analyze complex geometric data, such as complete intersection Calabi-Yau (CICY) manifolds \cite{32,33}. These studies demonstrate that NNs can identify physical regimes. Moreover, they detect boundaries between different classes of black holes,  predict regions of stability, and reveal hidden geometric features. Motivated by these successes, we apply machine learning models, including FCNNs \cite{34}, to identify whether a given black hole configuration exhibits Van der Waals behaviors based on  the  thermodynamic parameters.

The input data used in the present machine learning analysis are produced through  GPU-accelerated  simulations and stored in two separate files. The first file contains tuples of the form $(B,Q)$, where the parameters $(B,Q)$ describe the  black hole configurations investigated through the $P$-$V$ criticality condition given by  $\chi=\frac{3}{8}$. The second file also contains tuples $(B,Q)$, representing  stringy black hole configurations examined via the Joule--Thomson expansion with $\zeta=\frac{3}{4}$.
Accordingly, the dataset is divided into two classes, one related to the $P$-$V$ critical behavior and the other  one is linked  to the Joule--Thomson expansion. The labeling procedure is performed by comparing the parameter pairs $(B,Q)$ obtained from the two datasets. When a parameter pair appearing in the first dataset is also found in the second dataset, the corresponding sample receives the label $1$. Otherwise, the sample is assigned the label $0$. Concretely, we have the following scheme
\[
(B, Q) \rightarrow 0 \text{ or } 1.
\]
To ensure the quality of the datasets, we apply a filtering process to remove non-physical or redundant entries, retaining only parameter values within the allowed ranges. To improve convergence during trainings, all features are then normalized using standard scalings  to achieve a mean of zero and a variance of one.  The dataset is then divided into the  training, the  validation, and  the test subsets, ensuring that each black hole configuration appears in only one subset. This prevents any data leakage and allows the models to generalize effectively to unknown configurations. The resulting dataset is  used for the classification task. It is therefore well-suited for classification tasks, enabling machine learning models to reliably determine whether a deformed  black hole exhibits Van der Waals-like behaviors based on the input parameters.

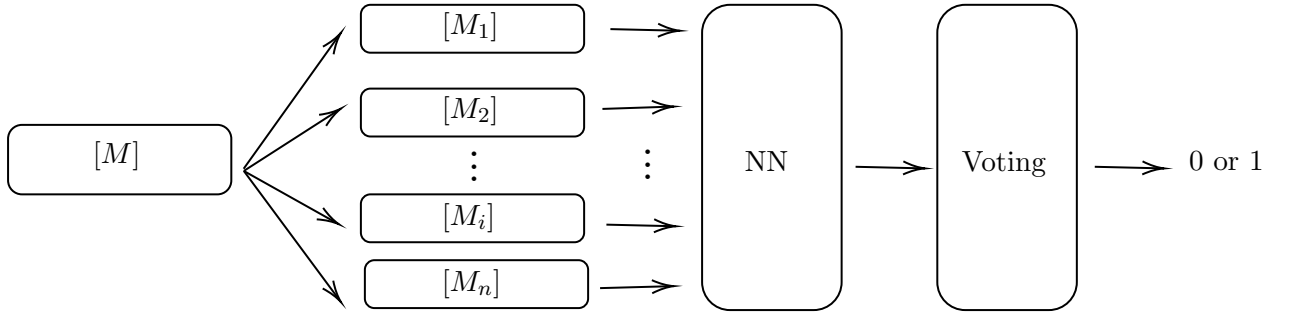
\begin{figure}[h!]
\centering
\begin{tikzpicture}[x=0.75pt,y=0.75pt,yscale=-1,xscale=1]

\draw (131.09,122.09) -- (178.63,186.21);
\draw [shift={(179.82,187.82)}, rotate=233.45, line width=0.75]
      (10.93,-3.29) .. controls (6.95,-1.4) and (3.31,-0.3) .. (0,0)
      .. controls (3.31,0.3) and (6.95,1.4) .. (10.93,3.29);

\draw (131.09,122.09) -- (177.07,147.84);
\draw [shift={(178.82,148.82)}, rotate=209.25, line width=0.75]
      (10.93,-3.29) .. controls (6.95,-1.4) and (3.31,-0.3) .. (0,0)
      .. controls (3.31,0.3) and (6.95,1.4) .. (10.93,3.29);

\draw (131.09,122.09) -- (178.14,91.9);
\draw [shift={(179.82,90.82)}, rotate=147.31, line width=0.75]
      (10.93,-3.29) .. controls (6.95,-1.4) and (3.31,-0.3) .. (0,0)
      .. controls (3.31,0.3) and (6.95,1.4) .. (10.93,3.29);

\draw (131,121) -- (178.65,54.44);
\draw [shift={(179.82,52.82)}, rotate=125.6, line width=0.75]
      (10.93,-3.29) .. controls (6.95,-1.4) and (3.31,-0.3) .. (0,0)
      .. controls (3.31,0.3) and (6.95,1.4) .. (10.93,3.29);

\draw (315,51) -- (347.82,51.77);
\draw [shift={(349.82,51.82)}, rotate=181.35, line width=0.75]
      (10.93,-3.29) .. controls (6.95,-1.4) and (3.31,-0.3) .. (0,0)
      .. controls (3.31,0.3) and (6.95,1.4) .. (10.93,3.29);

\draw (311,91) -- (344.82,89.88);
\draw [shift={(346.82,89.82)}, rotate=178.11, line width=0.75]
      (10.93,-3.29) .. controls (6.95,-1.4) and (3.31,-0.3) .. (0,0)
      .. controls (3.31,0.3) and (6.95,1.4) .. (10.93,3.29);

\draw (309.82,180.82) -- (344.82,179.87);
\draw [shift={(346.82,179.82)}, rotate=178.45, line width=0.75]
      (10.93,-3.29) .. controls (6.95,-1.4) and (3.31,-0.3) .. (0,0)
      .. controls (3.31,0.3) and (6.95,1.4) .. (10.93,3.29);

\draw (313,149) -- (345.82,149.77);
\draw [shift={(347.82,149.82)}, rotate=181.35, line width=0.75]
      (10.93,-3.29) .. controls (6.95,-1.4) and (3.31,-0.3) .. (0,0)
      .. controls (3.31,0.3) and (6.95,1.4) .. (10.93,3.29);

\draw (438,120) -- (470.82,120.77);
\draw [shift={(472.82,120.82)}, rotate=181.35, line width=0.75]
      (10.93,-3.29) .. controls (6.95,-1.4) and (3.31,-0.3) .. (0,0)
      .. controls (3.31,0.3) and (6.95,1.4) .. (10.93,3.29);

\draw (558,120) -- (590.82,120.77);
\draw [shift={(592.82,120.82)}, rotate=181.35, line width=0.75]
      (10.93,-3.29) .. controls (6.95,-1.4) and (3.31,-0.3) .. (0,0)
      .. controls (3.31,0.3) and (6.95,1.4) .. (10.93,3.29);

\draw (12.82,105.96) .. controls (12.82,102.12) and (15.94,99) .. (19.78,99) -- (117.85,99)
      .. controls (121.7,99) and (124.82,102.12) .. (124.82,105.96) -- (124.82,126.85)
      .. controls (124.82,130.7) and (121.7,133.82) .. (117.85,133.82) -- (19.78,133.82)
      .. controls (15.94,133.82) and (12.82,130.7) .. (12.82,126.85) -- cycle;

\draw (189.82,43.62) .. controls (189.82,40.97) and (191.97,38.82) .. (194.62,38.82)
      -- (297.02,38.82) .. controls (299.67,38.82) and (301.82,40.97) .. (301.82,43.62)
      -- (301.82,58.02) .. controls (301.82,60.67) and (299.67,62.82) .. (297.02,62.82)
      -- (194.62,62.82) .. controls (191.97,62.82) and (189.82,60.67) .. (189.82,58.02) -- cycle;

\draw (191.82,171.62) .. controls (191.82,168.97) and (193.97,166.82) .. (196.62,166.82)
      -- (299.02,166.82) .. controls (301.67,166.82) and (303.82,168.97) .. (303.82,171.62)
      -- (303.82,186.02) .. controls (303.82,188.67) and (301.67,190.82) .. (299.02,190.82)
      -- (196.62,190.82) .. controls (193.97,190.82) and (191.82,188.67) .. (191.82,186.02) -- cycle;

\draw (189.82,85.62) .. controls (189.82,82.97) and (191.97,80.82) .. (194.62,80.82)
      -- (297.02,80.82) .. controls (299.67,80.82) and (301.82,82.97) .. (301.82,85.62)
      -- (301.82,100.02) .. controls (301.82,102.67) and (299.67,104.82) .. (297.02,104.82)
      -- (194.62,104.82) .. controls (191.97,104.82) and (189.82,102.67) .. (189.82,100.02) -- cycle;

\draw (189.82,138.62) .. controls (189.82,135.97) and (191.97,133.82) .. (194.62,133.82)
      -- (297.02,133.82) .. controls (299.67,133.82) and (301.82,135.97) .. (301.82,138.62)
      -- (301.82,153.02) .. controls (301.82,155.67) and (299.67,157.82) .. (297.02,157.82)
      -- (194.62,157.82) .. controls (191.97,157.82) and (189.82,155.67) .. (189.82,153.02) -- cycle;

\draw (361,52.82) .. controls (361,45.09) and (367.27,38.82) .. (375,38.82)
      -- (417,38.82) .. controls (424.73,38.82) and (431,45.09) .. (431,52.82)
      -- (431,177.82) .. controls (431,185.55) and (424.73,191.82) .. (417,191.82)
      -- (375,191.82) .. controls (367.27,191.82) and (361,185.55) .. (361,177.82) -- cycle;

\draw (479,52.82) .. controls (479,45.09) and (485.27,38.82) .. (493,38.82)
      -- (535,38.82) .. controls (542.73,38.82) and (549,45.09) .. (549,52.82)
      -- (549,177.82) .. controls (549,185.55) and (542.73,191.82) .. (535,191.82)
      -- (493,191.82) .. controls (485.27,191.82) and (479,185.55) .. (479,177.82) -- cycle;

\draw (50,102) node [anchor=north west] { $\displaystyle [M]$};
\draw (225,37) node [anchor=north west] {$\displaystyle \left[M_1 \right]$};
\draw (225,79) node [anchor=north west] {$\displaystyle \left[M_2 \right]$};
\draw (225,132.82) node [anchor=north west] {$\displaystyle \left[M_i \right]$};
\draw (225,166) node [anchor=north west] {$\displaystyle \left[M_n \right]$};

\draw (239,107) node [anchor=north west] {\shortstack{\textbf{.}\\\textbf{.}\\\textbf{.}}};
\draw (326,106) node [anchor=north west] {\shortstack{\textbf{.}\\\textbf{.}\\\textbf{.}}};

\draw (377,107) node [anchor=north west] {NN};
\draw (486,107) node [anchor=north west] {Voting};
\draw (600,107) node [anchor=north west] {0 or 1};

\end{tikzpicture}
\caption{Voting procedure for Van der Waals behaviors of $B$-deformed RN-AdS  black holes.} \label{5}
\end{figure}

To train the model, we utilize a feed-forward NN that inputs the tuples and predicts the corresponding label 0 or 1. In order to improve accuracy and reduce sensitivity features  to small variations in the  input parameters, we adopt a voting strategy for each input tuple
$
[M] = (B, Q),
$ 
where multiple equivalent representations 
$
[M_1, M_2, \dots, M_n]
$ 
are generated. Each representation is handled independently by the NN techniques. The final prediction is determined by a majority vote among all outputs. This approach, illustrated in Fig. (\ref{5}), enhances the classification performance. In addition, it improves a  generalization to new black hole configurations.

As depicted in Fig. (\ref{5}), each input tuple is first fed into the NN  producing  predictions for all its variants. Such  individual predictions are then aggregated using a voting algorithm to determine the final output label, indicating whether the black hole exhibits Van der Waals-type behaviors. This combination of  the rigorous data preparation and  the advanced learning strategy ensure an accurate and a reliable classification. Now,  we move to  investigate the Van der Waals behavior of stringy  black hole configurations using a NN constructed from fully connected layers. The dataset comprises 467,568 samples being characterized by  the couple  $(B, Q)$. To ensure a balanced dataset, we select an equal number of  Van der Waals and non--Van der Waals samples. After shuffling, the data are split into 70\% training, 20\% validation, and 10\% testing subsets. To classify the Van der Waals behaviors from the 2-dimensional input vectors, we employ a FCNN with a two-class probability output. The network architecture is defined as
\[
N_{2 \times 2} = F \circ G_{2 \times 32}, \; F \circ G_{32 \times 64}, \; F \circ G_{64 \times 32}, \; S \circ G_{32 \times 2},
\]
where $G_{n \times m}$ denotes a fully connected layer with $n$ inputs and $m$ outputs,  $F$   is the
standard Relu function, BatchNorm normalization and dropout procedure, and $S$ is a softmax layer producing the output probabilities. The algorithmic model is trained on the training dataset using the  binary cross-entropy as the loss function and the Adam optimizer. During training, the learning rate is automatically adjusted to guarantee a smooth convergence of the model.

\begin{figure}[h!]
\centering
\includegraphics[width=0.6\textwidth]{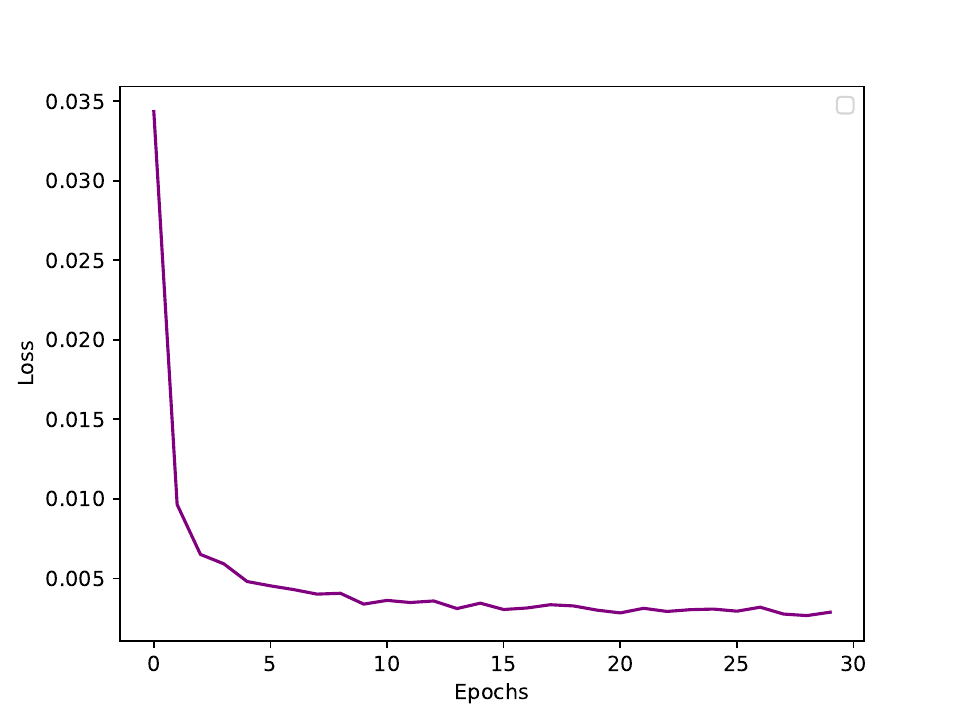} 
 \caption{Training curves for the FCNN model.} \label{6}
\end{figure}

As shown in Fig.(\ref{6}), the loss decreases rapidly during the first few epochs and then stabilizes. This indicates that the network is learning efficiently and converging quickly. Both training and validation accuracies remain above 99\%. This demonstrates that the model generalizes well without overfitting. Slight fluctuations in  the validation loss are expected due to random variations within batches and do not affect the l downward trend. Evaluations on the test set yield a test accuracy of the following value
\[
\text{Test Accuracy} = 0.9918.
\]
This  reveals that the network can achieves classification accuracy distinguishing  the Van der Waals behaviors  from non–Van der Waals ones. Such a high accuracy indicates that the model has learned a clear and stable boundary between the two regimes in the 2-dimensional input space. The performance metrics are summarized in Table~\ref{tab1}.
\begin{table}[!ht]
 \centering 
 \caption{Model performance with FCNN.}
 \label{tab1} 
 \begin{tabular}{l S[round-precision=5]}
   \toprule
    \textbf{Metric} & {\textbf{Van der Waals}} \\ 
    \midrule 
    \textbf{Test Accuracy} &  0.9918 \\ 
    \textbf{Voting} & 1 \\ 
    \bottomrule
 \end{tabular}
\end{table}
In order to evaluate the robustness of the classifier in more detail, we employ a voting system inspired by methods commonly used in a symmetry classification. For each configuration, 100 perturbed samples preserving symmetries are independently generated and classified. The final label is assigned by a majority vote, making it possible to measure the stability of the model under structured perturbations. This procedure provides a voting accuracy of the following value 
\[
\text{Voting Accuracy} = 1.
\]
This demonstrates  that the classifier remains fully consistent under small perturbations and generalizes strongly beyond the exact training distribution. The confusion matrix in Table~\ref{tab2} confirms that the model correctly classifies almost all samples into the two categories, with only a negligible number of the  classification errors. This shows the high reliability of the classifier and its ability to capture the essential thermodynamic distinctions between Van der Waals and non-Van der Waals behaviors.

\begin{table}[ht]
\centering
\caption{Confusion matrix with FCNN.}
\label{tab2}
\begin{tabular}{lcc}
\toprule
\textbf{A/P} & \textbf{Class 0} & \textbf{Class 1} \\
\midrule
\textbf{Class 0} & 92,705 & 37 \\
\textbf{Class 1} & 92 & 680 \\
\bottomrule
\end{tabular}
\end{table}

\section{Machine learning applied to   optics of  $B$-deformed  RN-AdS black holes}
In this section, we explore the optical properties of the proposed black holes using machine learning methods. We begin by analyzing the characteristics of their shadows. We then use observations reported by  the EHT international collaborations  to constrain the black hole parameters including the stringy one. Finally, we employ machine learning techniques to investigate the relationship between the EHT observational data and the optical signatures of these black holes.
\subsection{Shadow behaviors  of non-rotating  solutions}
In this part, we  approach  the optical properties of these stringy  charged black holes by  examining  the  geometric structure of their shadows. To do so, we first discuss the motion of photons around such  black holes via the  Euler-Lagrange equations
\begin{equation}
\frac{d}{d\sigma}\left(\frac{\partial \mathcal{L}}{\partial \dot{x}^{\mu}}\right)
=
\frac{\partial \mathcal{L}}{\partial x^{\mu}},
\label{eq:4}
\end{equation}
where $\sigma$ is the affine parameter and  $\dot{x}^{\mu}$ represents the photon  four-velocity.  $\mathcal{L}$ is the Lagrangian density of the photon system given by 
\begin{equation}
\mathcal{L}
=
-\frac{1}{2} g_{\mu\nu}
\frac{dx^{\mu}}{d\sigma}
\frac{dx^{\nu}}{d\sigma}
\label{eq:5}
\end{equation}
Using the metric form, this reduces to 
\begin{equation}
\mathcal{L}
=
\frac{1}{2}
\left(
f(r)\dot{t}^{\,2}
-
\frac{\dot{r}^{\,2}}{f(r)}
-
r^{2}
\left(
\dot{\theta}^{\,2}
+
\sin^{2}\theta \, \dot{\phi}^{\,2}
\right)
\right).
\label{eq:55}
\end{equation}

In what follows, we consider  a restriction   motion of the photon to the plane of the equator  by taking  $\theta=\pi/2$ and $\dot{\theta}=0$. In this way,  the metric dependence on the coordinates $t$ and $\phi$  can be reduced. This provides two  the conserved quantities being  the  energy $E$ and  the angular momentum $L$ expressed as follows 
\begin{eqnarray}\
E
=
f(r)\frac{dt}{d\sigma}, \qquad 
L
=
r^{2}\frac{d\phi}{d\sigma}.
\label{eq:8}
\end{eqnarray}\
The effective potential of the radiation motion of the photon can be obtained  via  the normalization condition of the null geodesic $
g_{\mu\nu}\dot{x}^{\mu}\dot{x}^{\nu}=0$ which takes the following form 
\begin{equation}
\left(\frac{dr}{d\sigma}\right)^2+V_{\mathrm{eff}}(r)=0.
\label{eq:9}
\end{equation}
In this way,  the effective potential  can be written as
\begin{equation}
V_{\mathrm{eff}}(r)
=
f(r)
\left(
\frac{L^{2}}{r^{2}}
-
\frac{E^{2}}{f(r)}
\right).
\label{eq:10}
\end{equation}
Combining these equations,  the orbital equation describing the motion of the photon  can be obtained by approaching the relation 
\begin{equation}
\frac{dr}{d\phi}
=
\pm r
\sqrt{
f(r)
\left(
\frac{r^{2}E^{2}}{L^{2}f(r)}
-
1
\right)
}.
\label{eq:11}
\end{equation}
The critical points can be derived  by solving the following constraints
\begin{equation}
V_{\mathrm{eff}}(r_{p})=0,
\qquad
\left.
\frac{dV_{\mathrm{eff}}(r)}{dr}
\right|_{r=r_{p}}
=0.
\label{eq:12}
\end{equation}
Taking into account the critical point in the orbital motion of the photon $
\left.
\frac{dr}{d\phi}
\right|_{r=r_{p}}
=0$, 
one gets 
\begin{equation}
\frac{dr}{d\phi}
=
\pm r
\sqrt{
f(r)
\left(
\frac{r^{2}f(r_{p})}{r_{p}^{2}f(r)}
-
1
\right)
}.
\label{eq:14}
\end{equation}
 To handle these equations,  one  could follow the formalism developed in  ~\cite{61,62,63}.   Indeed, 
a static observer placed at $r_0$ sends a light ray into the past forming an  angle $\beta$ corresponding to the radial direction expressed as 
\begin{equation}
\cot\beta
=
\left.
\frac{\sqrt{g_{rr}}}{\sqrt{g_{\phi\phi}}}
\frac{dr}{d\phi}
\right|_{r=r_{0}}
\label{eq:15}
\end{equation}
leading to 
\begin{equation}
\sin^{2}\beta
=
\frac{f(r_{0})\,r_{p}^{2}}
{r_{0}^{2}f(r_{p})}.
\label{eq:16}
\end{equation}
In the limit where $r$ approaches the photon sphere radius $r_{p}$, the shadow radius measured by a static observer located at $r_{0}$ takes the form
\begin{equation}
r_{s}
=
r_{0}\sin\beta.
\label{eq:17}
\end{equation}
The computations give
\begin{equation}
r_{s}
=
r_{p}
\sqrt{
\frac{f(r_{0})}{f(r_{p})}
}.
\label{eq:17}
\end{equation}
Using Eqs.~(\ref{eq:4}) and (\ref{eq:12}), and the  condition $V_{\mathrm{eff}}'=0$, the radius of the photon  sphere associated with the stringy  charged AdS black holes holds the following equation 
\begin{equation}
2Q^{2}
-
3Mr_{p}
+
r_{p}^{2}
-
\frac{(B+\Lambda) r_{p}^{4}}{3} 
=
0.
\label{eq:18}
\end{equation}
It has been observed that the determination of the  analytic  solutions is  a tough task. However, the photon sphere radius  can be  obtained numerically by solving the corresponding constraint equation. For a static observer situated at spatial infinity, one has $f(r_{0})=1$, which enables the computation  of the black hole shadow radius.

To provide a clearer visualization of the results, we assess the   shadow optical behaviors.   In the space of the observer, namely in the plane perpendicular to the line of sight, the shadow shape  can be  described through the celestial coordinates. For an observer located far away from the black hole, these coordinates are defined as follows

\begin{equation}
X = \lim_{r \to \infty} \left( -r^2 \sin \theta_0 \frac{d\phi}{dr} \right)_{\theta_0 \to \frac{\pi}{2}}, \qquad
Y = \lim_{r \to \infty} \left( r^2 \frac{d\theta}{dr} \right)_{\theta_0 \to \frac{\pi}{2}},
\end{equation}
where $(r_0,\theta_0)$ are the position coordinates of the observer. It  is denoted that  the  celestial coordinate $X$ represents the observed perpendicular distance of the image, and $Y$ indicates  the perpendicular distance projected onto the equatorial plane.

\begin{figure}[th]
\centering
\includegraphics[scale=0.3]{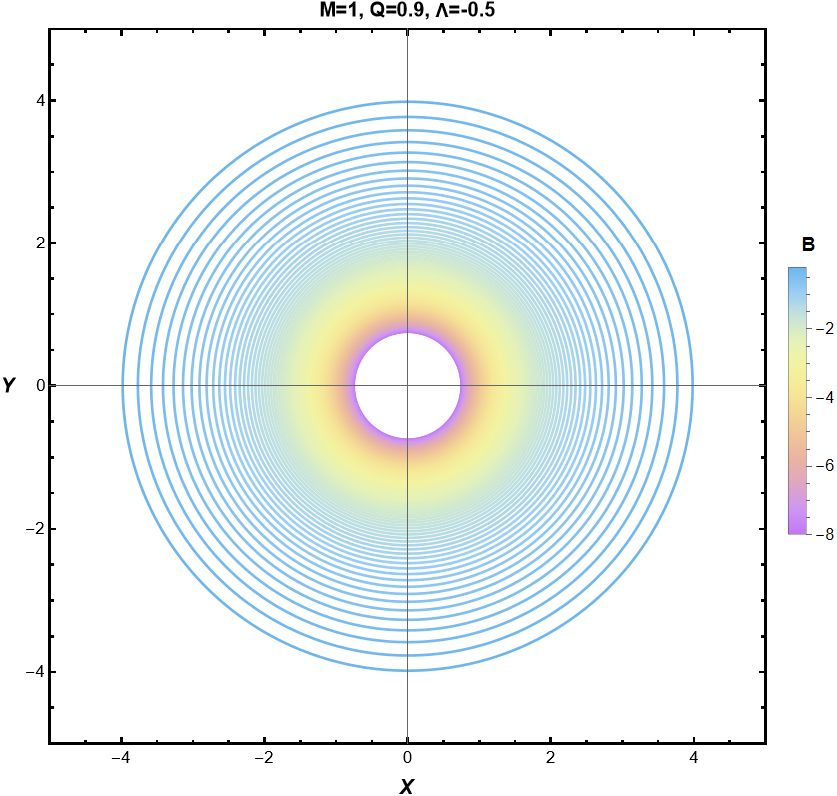} \includegraphics[scale=0.3]{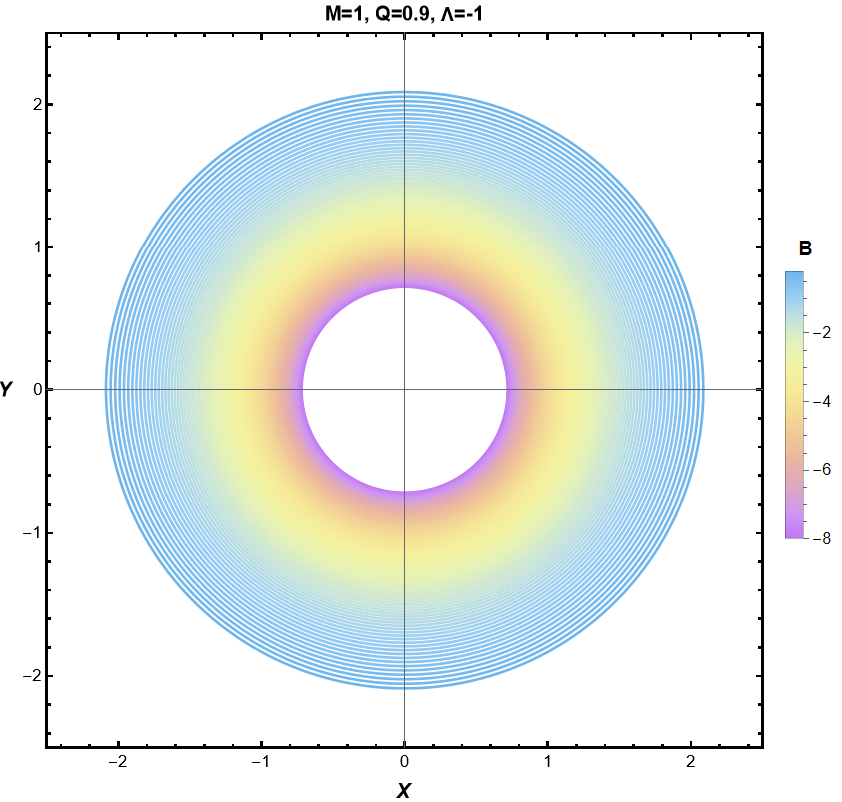}
\includegraphics[scale=0.3]{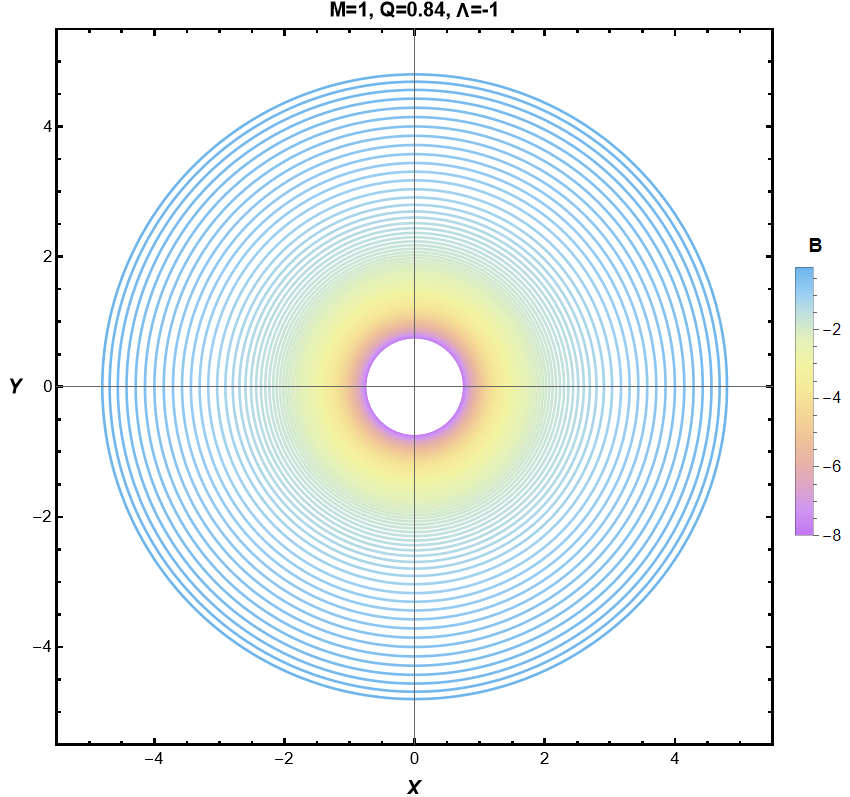}
\caption{\textit{\protect\footnotesize Effect of the  $B$ parameter on shadow curves.}}
\label{sh2}
\end{figure}

\begin{figure}[h!]
\centering
\includegraphics[scale=0.29]{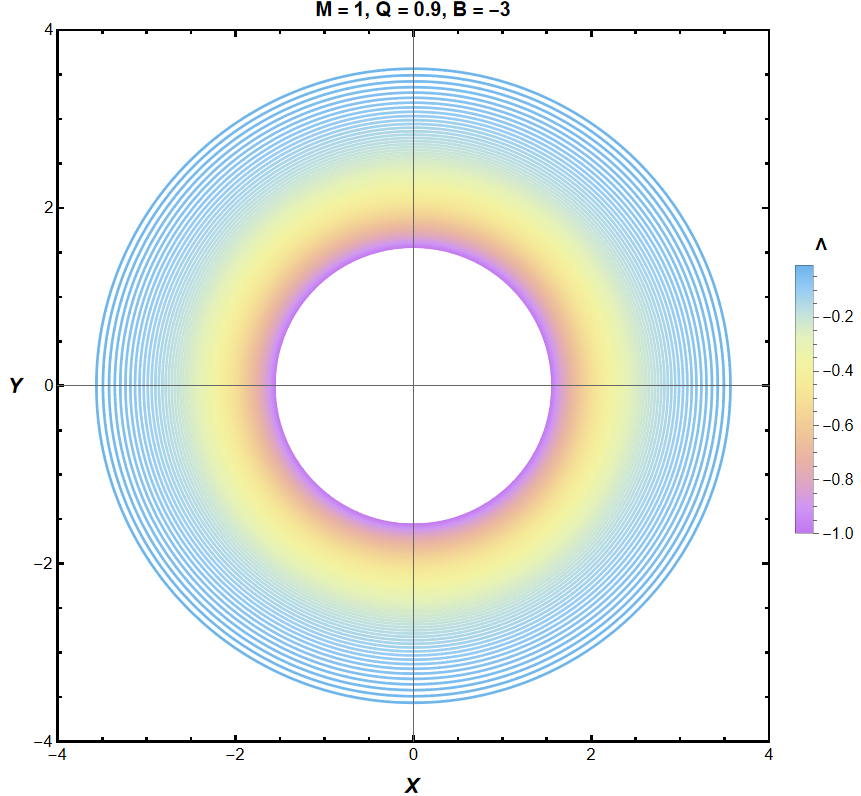} \includegraphics[scale=0.29]{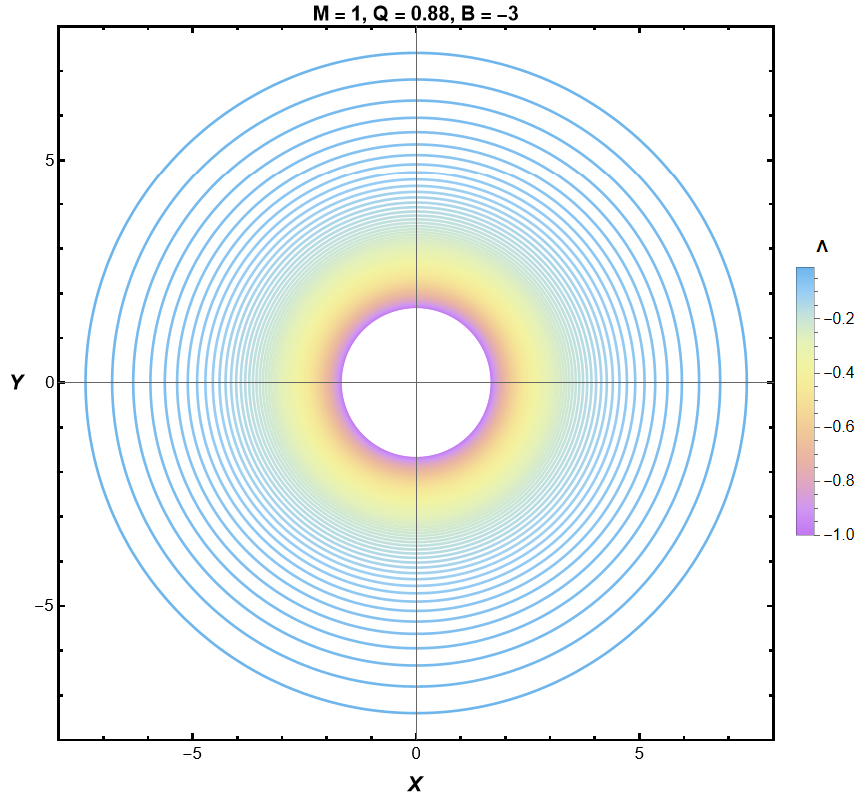}
\includegraphics[scale=0.3]{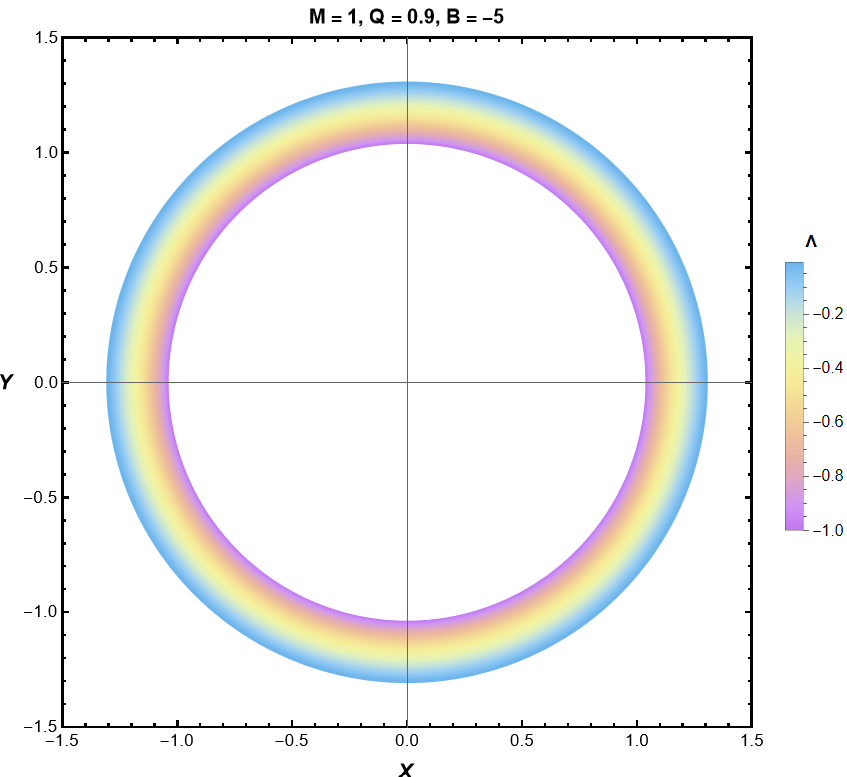}
\caption{\textit{\protect\footnotesize Effect of $\Lambda$ parameter on shadow curves.}}
\label{sh33}
\end{figure}

We now examine the effects of the stringy parameter $B$, the cosmological constant $\Lambda$, and the charge parameter $Q$ on the shadow behavior. As shown in Fig.~(\ref{sh2}), this stringy  parameter $B$ controls the size of the shadow.  Decreasing $B$ leads to a larger and more extended shadows.
 A similar behavior is observed in Fig.~(\ref{sh33}) for the cosmological constant, since increasing $\Lambda$ also enlarges the shadow radius.
 
\begin{figure}[th]
\centering
\includegraphics[scale=0.3]{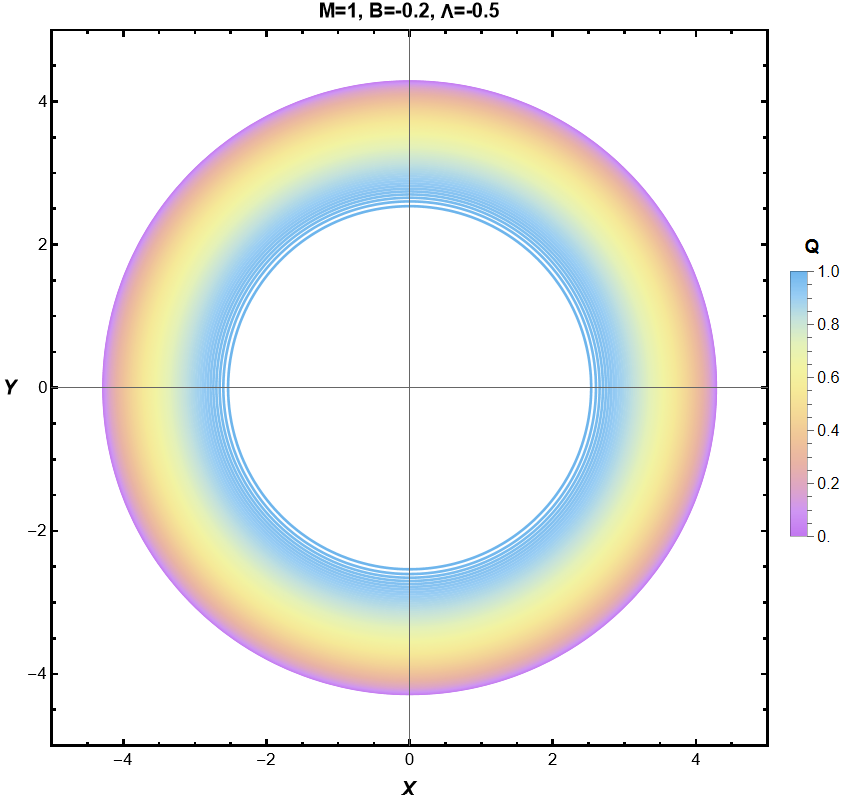} \includegraphics[scale=0.3]{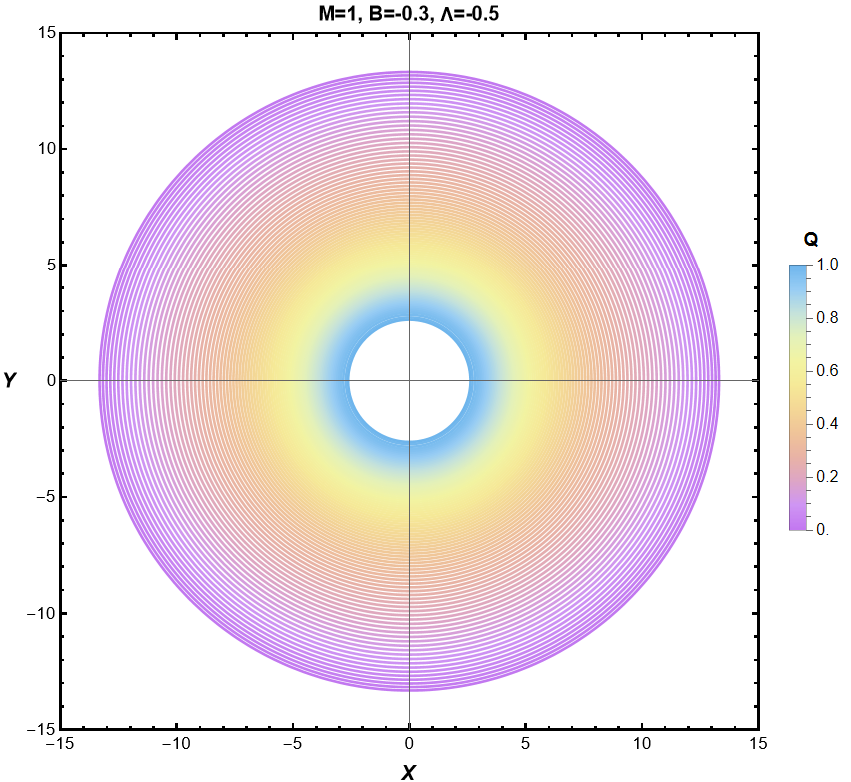}
\includegraphics[scale=0.3]{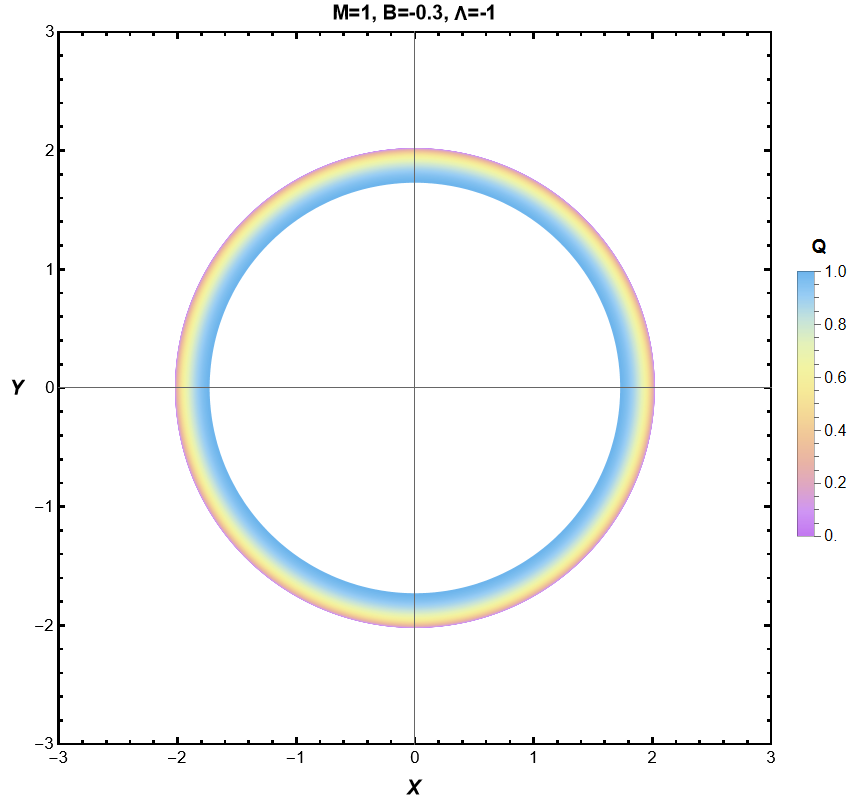}
\caption{\textit{\protect\footnotesize Effect of $Q$ parameter on shadow curves.}}
\label{sh1}
\end{figure}

  In contrast, the charge parameter exhibits the opposite effect, as increasing the charge reduces the size of the shadow as shown in Fig. (\ref{sh1}). 
  
  However, despite their influence on the shadow radius, these three parameters do not affect globally  the shape of the shadow, which remains circular since we are considering non-rotating solutions.

\subsection{ Parameter constrining scenarios via  EHT observations}
In order to provide a connection between theoretical predictions and empirical data, this  part   exposes  an analysis of the shadow of  non-rotating  stringy AdS black holes, in comparison with the observational results reported by the EHT international collaboration \cite{cc1c}.  Precisely, we  constrain the parameters of such  black hole models using the  observational data associated  with  the black holes M87* and Sgr A* ~\cite{AE1,AE2,AE3}. In practice, these constrained values can be obtained from the difference in the diameter of the black hole shadow relative to that of a Schwarzschild black hole, defined as follows
\begin{equation}
{\ d} = \frac{R_s}{r_{sh}}-1,
\end{equation}
where $R_s$ denotes the shadow radius and $r_{\text{sh}}$ represents the Schwarzschild radius. The dimensionless ratio $R_s/M$ serves as a primordial observable  to  confront the  rational models with empirical  data.  The  $1-\sigma$  and  $2-\sigma$  confidence intervals derived from the EHT observations are displayed in Table ~\ref{t1}.
\begin{table}[h!]
\centering
\begin{tabular}{|c|c|c|c|}
\hline
\textbf{Black hole} & \textbf{Deviation ($d$)} & \textbf{1-$\sigma$ bounds}
& \textbf{2-$\sigma$ bounds} \\ \hline
M87$^*$ (EHT) & $-0.01^{+0.17}_{-0.17}$ & $4.26 \leq \frac{R_s}{M} \leq 6.03$
& $3.38 \leq \frac{R_s}{M} \leq 6.91$ \\ \hline
Sgr~A$^*$ (EHT$_{\text{VLTI}}$) & $-0.08^{+0.09}_{-0.09}$ & $4.31 \leq \frac{%
R_s}{M} \leq 5.25$ & $3.85 \leq \frac{R_s}{M} \leq 5.72$ \\ \hline
Sgr~A$^*$ (EHT$_{\text{Keck}}$) & $-0.04^{+0.09}_{-0.10}$ & $4.47 \leq \frac{%
R_s}{M} \leq 5.46$ & $3.95 \leq \frac{R_s}{M} \leq 5.92$ \\ \hline
\end{tabular}%
\caption{ \textit{\protect\footnotesize  Fractional deviations and
corresponding bounds for M87$^*$ and Sgr~A$^*$ black holes.}}
\label{t1}
\end{table}

\begin{figure}[h!]
\centering
\includegraphics[scale=0.34]{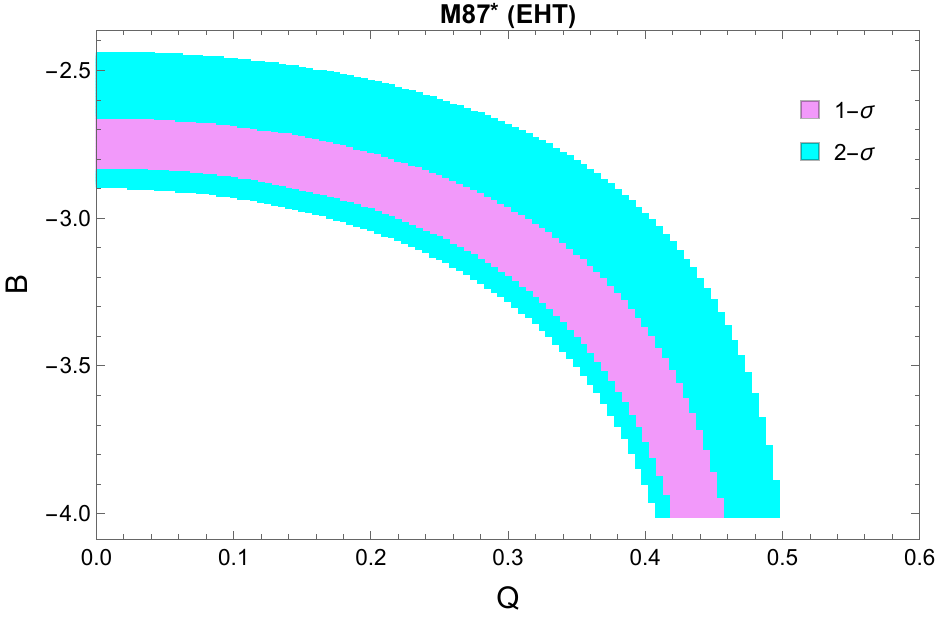} \includegraphics[scale=0.34]{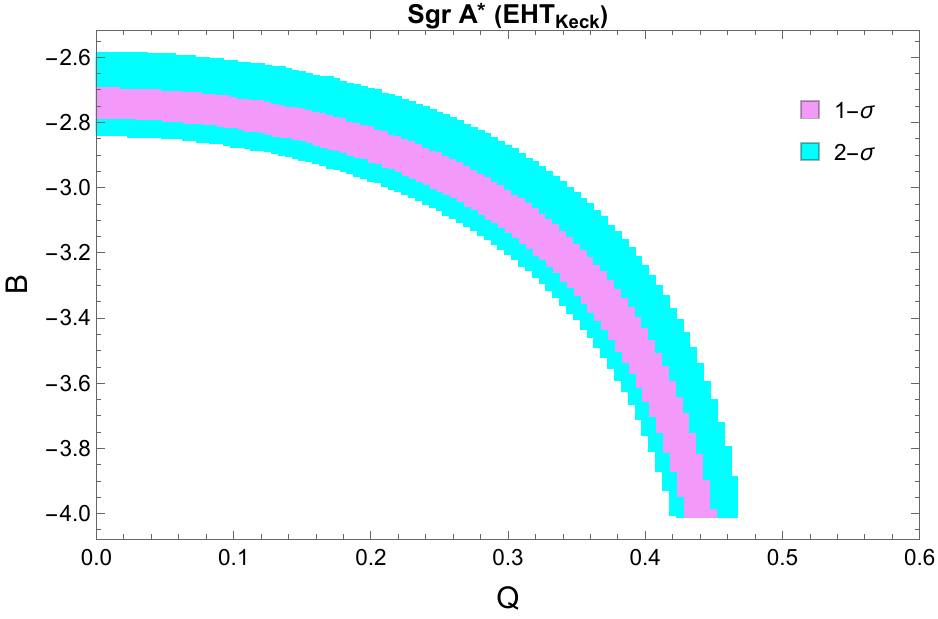}
\includegraphics[scale=0.34]{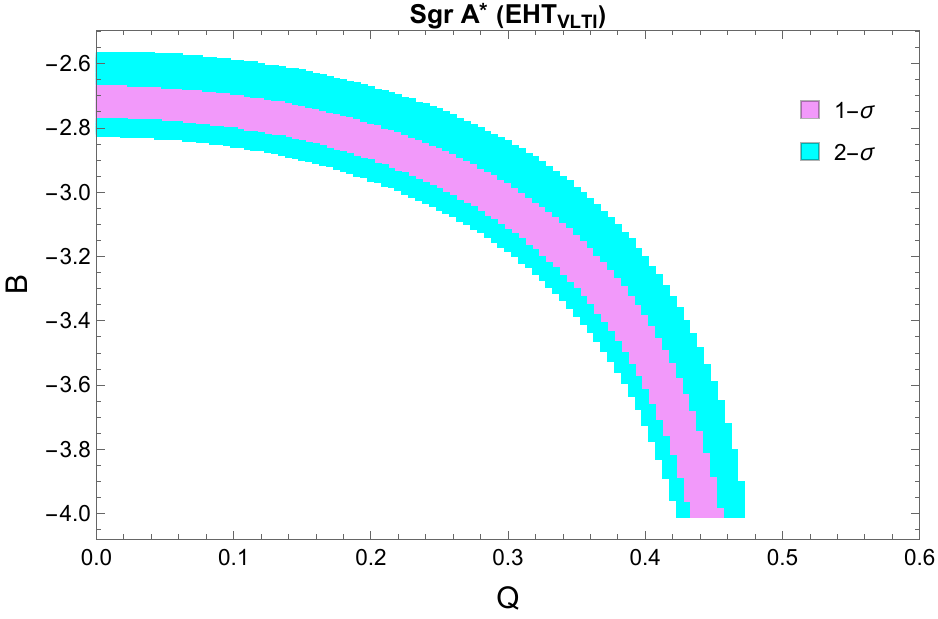}
\caption{\textit{\protect\footnotesize Constraint regions in the $(B, Q)$ plane with $M=1$ and  $\Lambda=-2.5
$.}}
\label{sh3}
\end{figure}

In the following discussions, we elaborate a numerical procedure  computations to determine the parameter pair $(B, Q)$ that yield black hole shadow configurations consistent with  EHT observational data.   In particular,  we consider  fixed values $\Lambda=-2.5$ and $M=1$. The corresponding
results are shown in  Fig.~(\ref{sh3}).

As illustrated in the three panels of this figure, the density of points consistent with the
empirical observations increases for higher values of  $Q$. This result suggests that the
space-time background of the black hole can effectively reproduce observational signatures.
Given the interdependence between the two parameters, we fix one parameter while constraining the remaining ones. For $Q=0.3$, Table (\ref{s1}) presents the corresponding ranges of $B$  matching with  $1-\sigma$ and $2-\sigma$ confidence intervals for the three experimental scenarios of the  black hole models.

\begin{table}[!ht]
\centering
\caption{Constraints on the parameter $B$ for  $Q=0.3$, derived from observational bounds at the $1-\sigma$ and $2-\sigma$ confidence levels.}
\label{tab11}
\begin{tabular}{llccc}
\toprule
 & \textbf{Confidence} & \textbf{$M87^*$} & \textbf{$SgrA^*_{\mathrm{Keck}}$} & \textbf{$SgrA^*_{\mathrm{VLTI}}$} \\
\midrule

\multirow{2}{*}{$ $}
& $1-\sigma$ & $-2.97 < B < -2.65$ & $-2.93< B < -2.69$ &$-2.9 < B < -2.66$ \\
& $2-\sigma$ & $-3.03 < B < -2.45$ & $-2.97 < B < -2.59$ & $-2.95 < B < -2.57$ \\
\bottomrule
\end{tabular} \label{s1}
\end{table}
A clear conclusion emerges from this examination. Indeed, although the moduli  space theoretically allows for large regions, only limited regions are compatible with the observational constraints. In particular, the   stringy parameter $B$ plays a major role, as it is consistently constrained by the data in all cases, whereas the remaining  parameters primarily influence the shape and  the size of the allowed regions. This indicates that $B$ behaves as a control parameter  controlling  the matching  between the theoretical model and the observed black hole shadows.

\subsection{Machine learning constraints on  stringy  black hole parameters from EHT observations}
\begin{figure}[!h]
\centering
\includegraphics[scale=0.32]{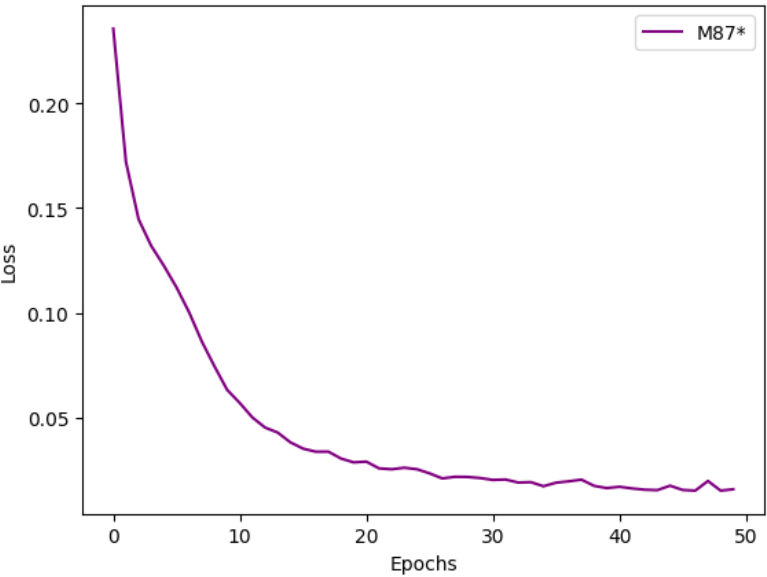}\hspace{2mm}
\includegraphics[scale=0.32]{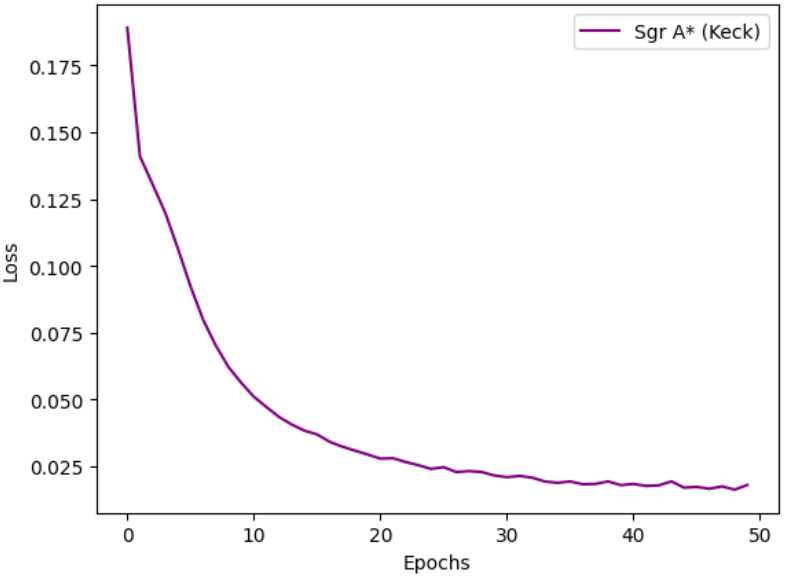}\hspace{2mm}
\includegraphics[scale=0.32]{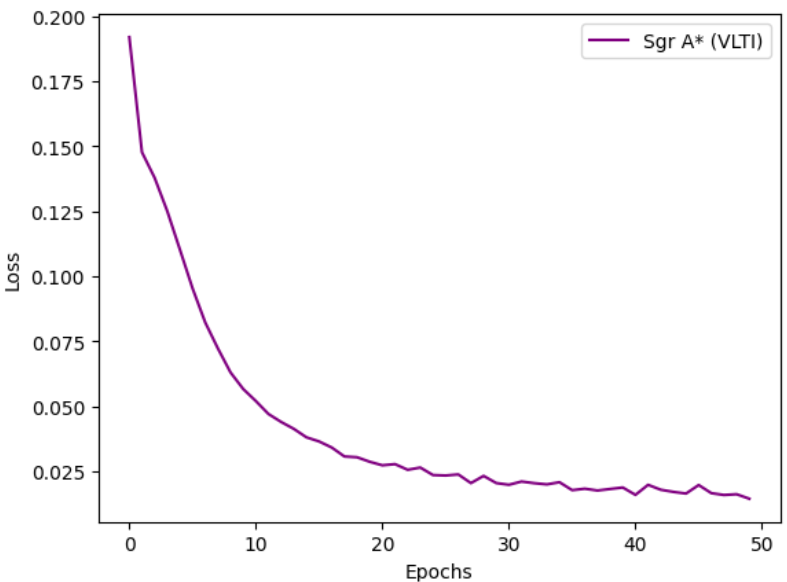}
\caption{Training curves of the FCNN model for the $1-\sigma$ observational constraint.}
\label{fig:train}
\end{figure}

\begin{figure}[!h]
\centering
\includegraphics[scale=0.32]{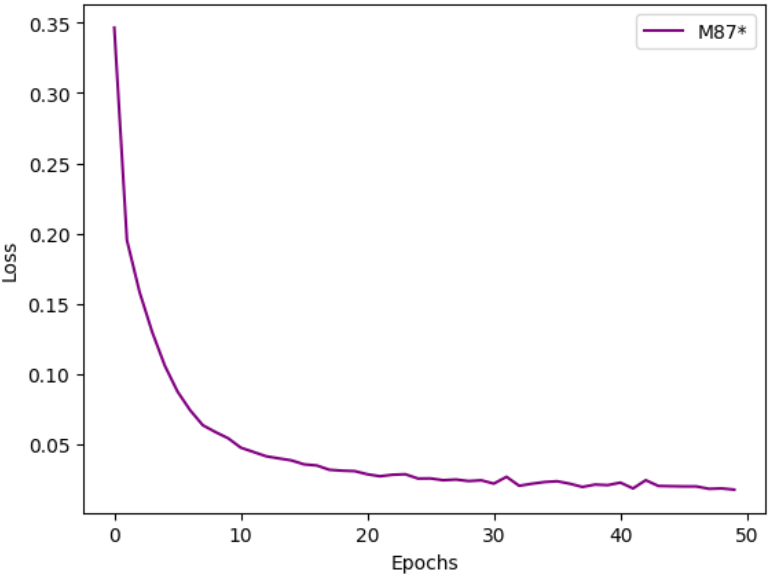}\hspace{1mm}
\includegraphics[scale=0.32]{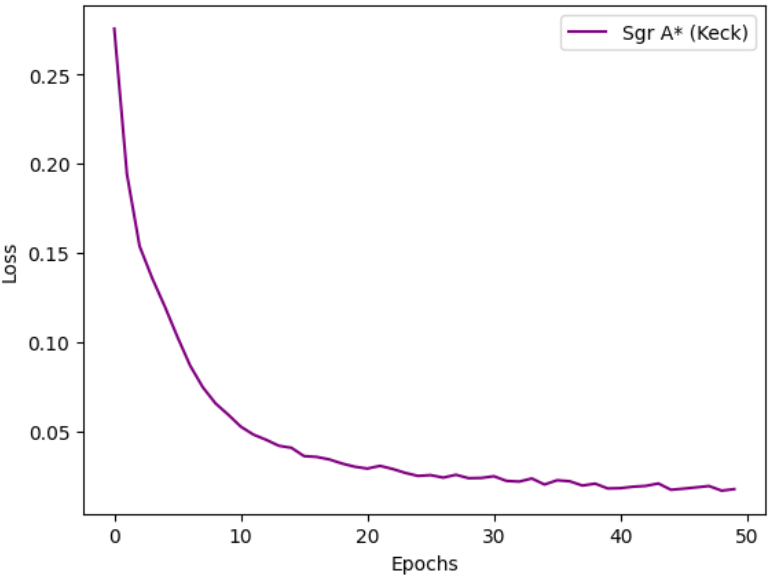}\hspace{1mm}
\includegraphics[scale=0.32]{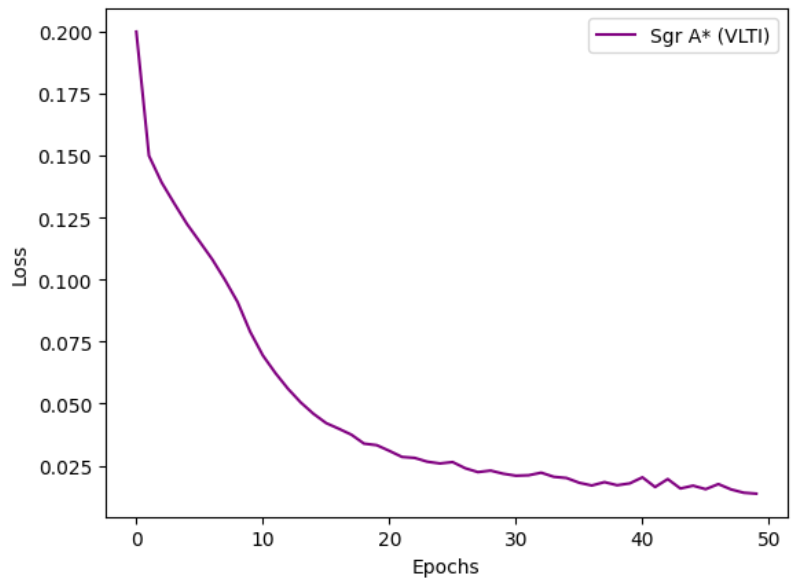}
\caption{Training curves of the FCNN model for the $2-\sigma$ observational constraint.}
\label{fig:train1}
\end{figure}

Now we move to  employ machine learning techniques to analyze stringy  black hole shadow configurations. After that, we  examine their consistency with the observational constraints provided by the EHT collaboration. Such an approach is particularly effective for exploring extended regions of the parameter space and identifying physically viable configurations compatible with the observational data \cite{y1}. Following the same methodology adopted previously in the thermodynamic analysis, the datasets are generated through numerical simulations by scanning different combinations of the parameters $(B,Q)$. For each configuration, the corresponding black hole shadow is evaluated and confronted with the observational bounds associated with the cases of M87*, SgrA*$_{\mathrm{VLTI}}$, and SgrA*$_{\mathrm{Keck}}$ models, considering both the $1-\sigma$ and $2-\sigma$ confidence levels. Each configuration is then assigned a binary label according to its agreement with the EHT constraints according the  scheme 
\begin{equation}
(B,Q)\rightarrow 0 \ \text{or}\ 1,
\end{equation}
where 1 corresponds to an observationally allowed configuration, while 0 denotes an excluded one.\\
\begin{figure}[!ht]
\centering
\includegraphics[scale=0.26]{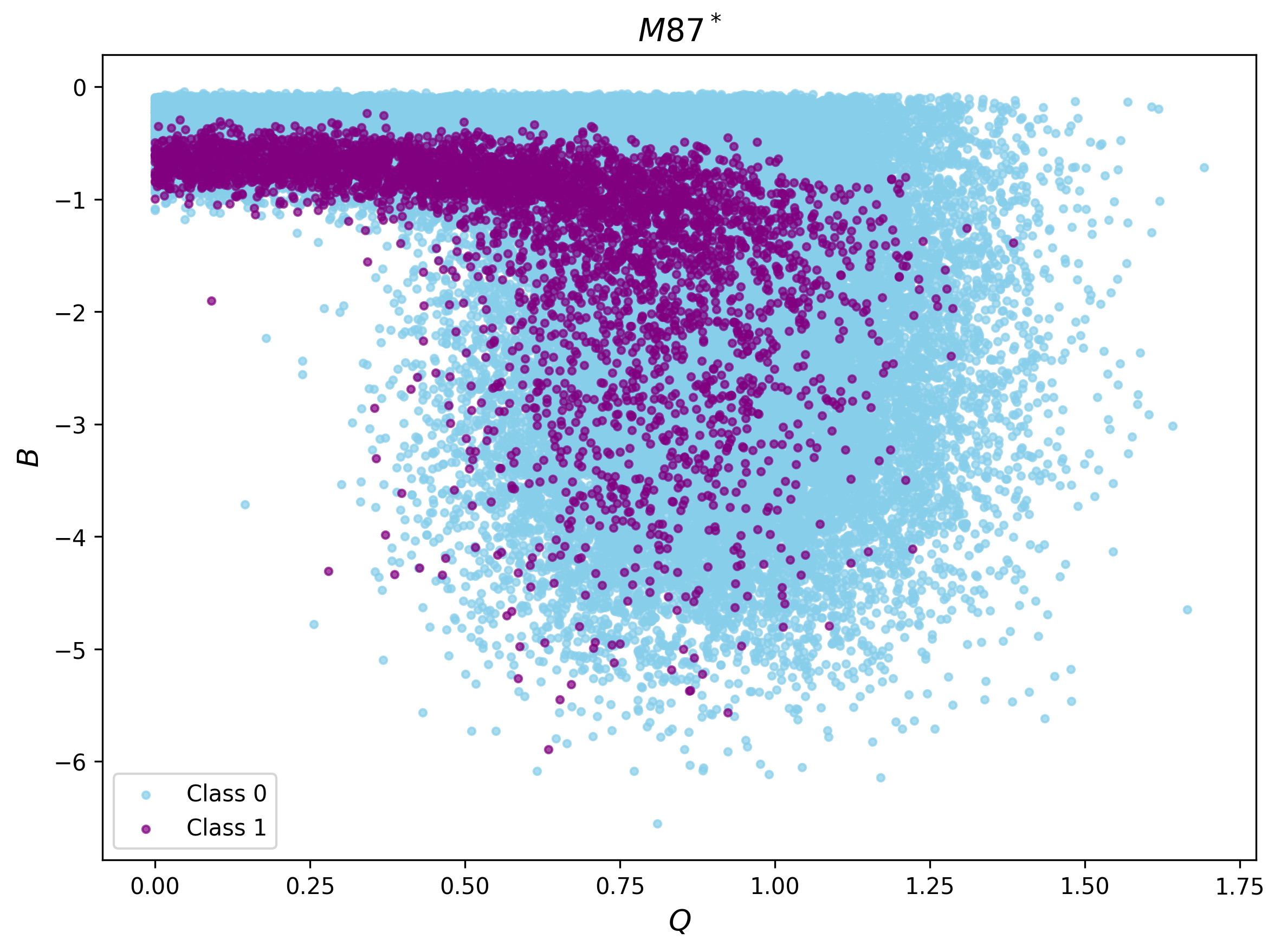}\hspace{2mm}
\includegraphics[scale=0.26]{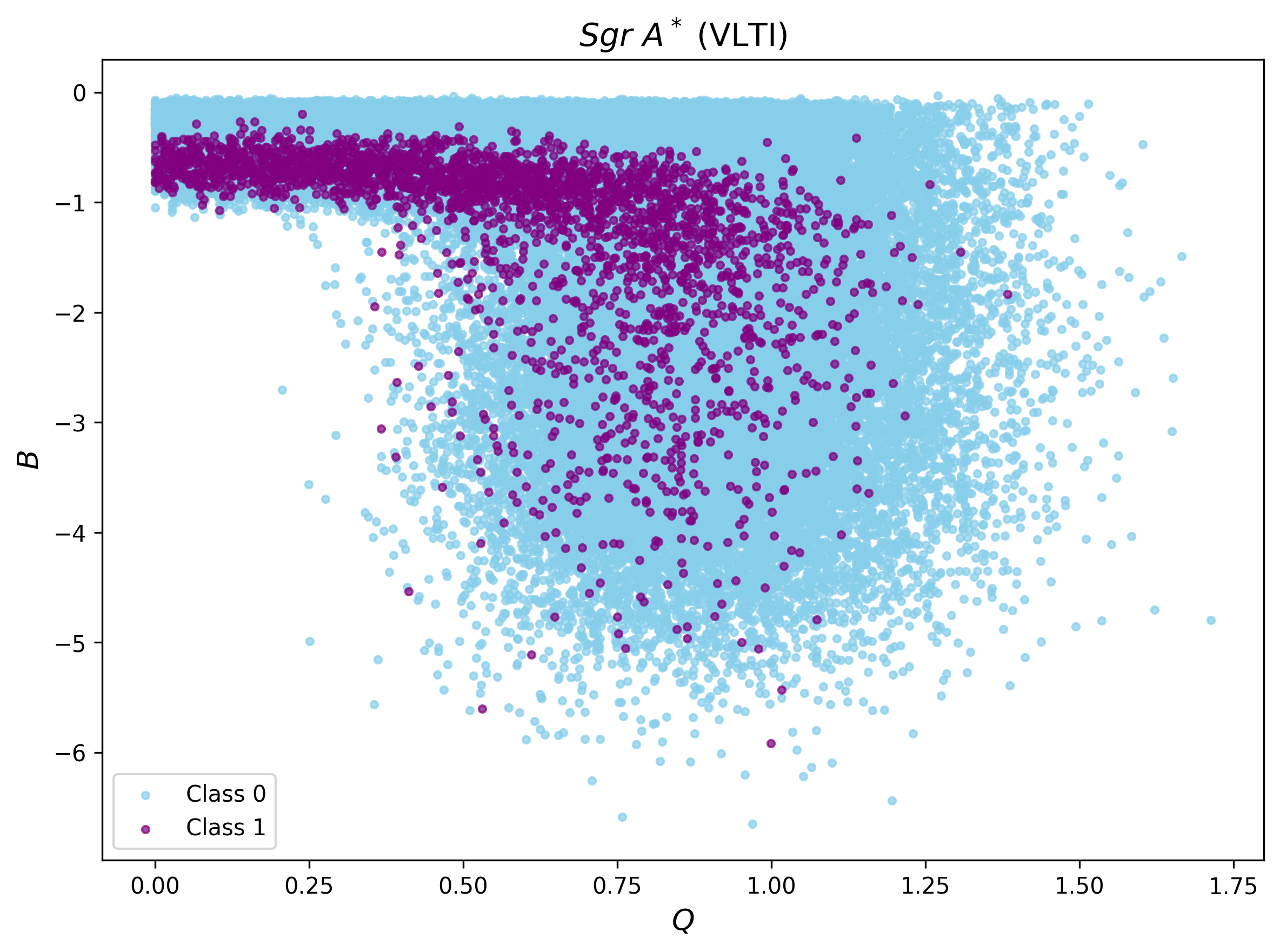}\hspace{2mm}
\includegraphics[scale=0.26]{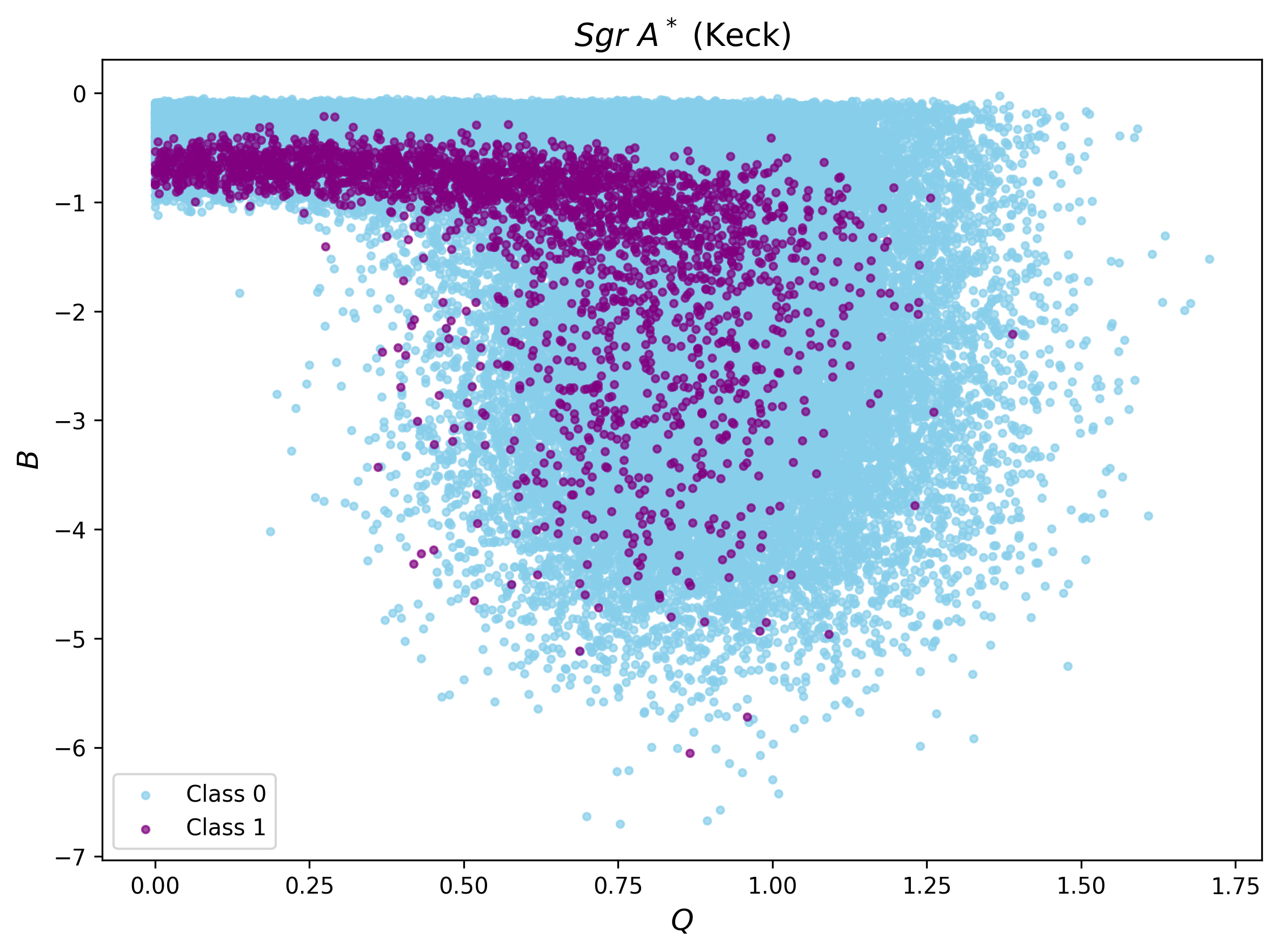}
\vspace{2mm}
\includegraphics[scale=0.26]{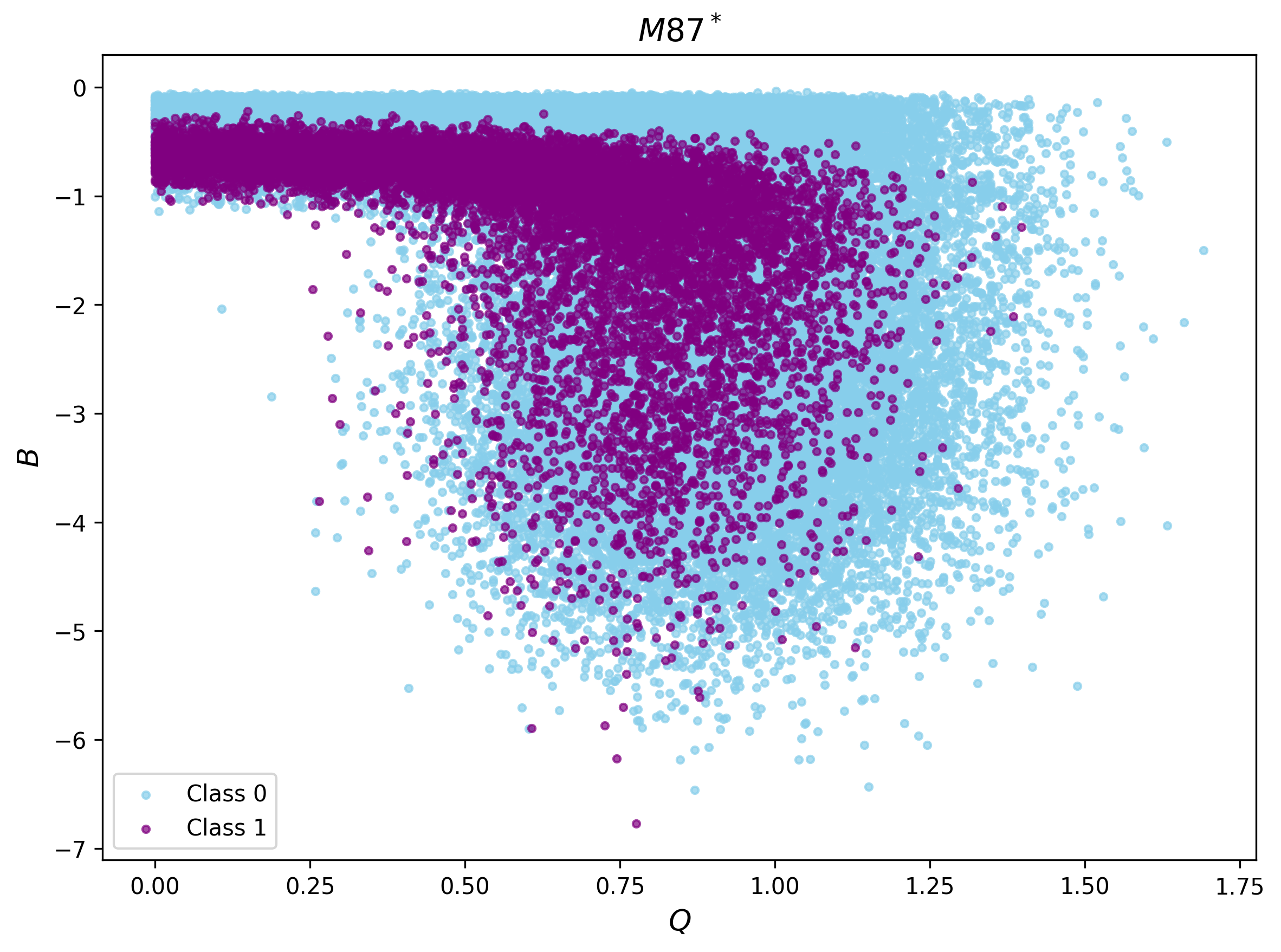}\hspace{2mm}
\includegraphics[scale=0.26]{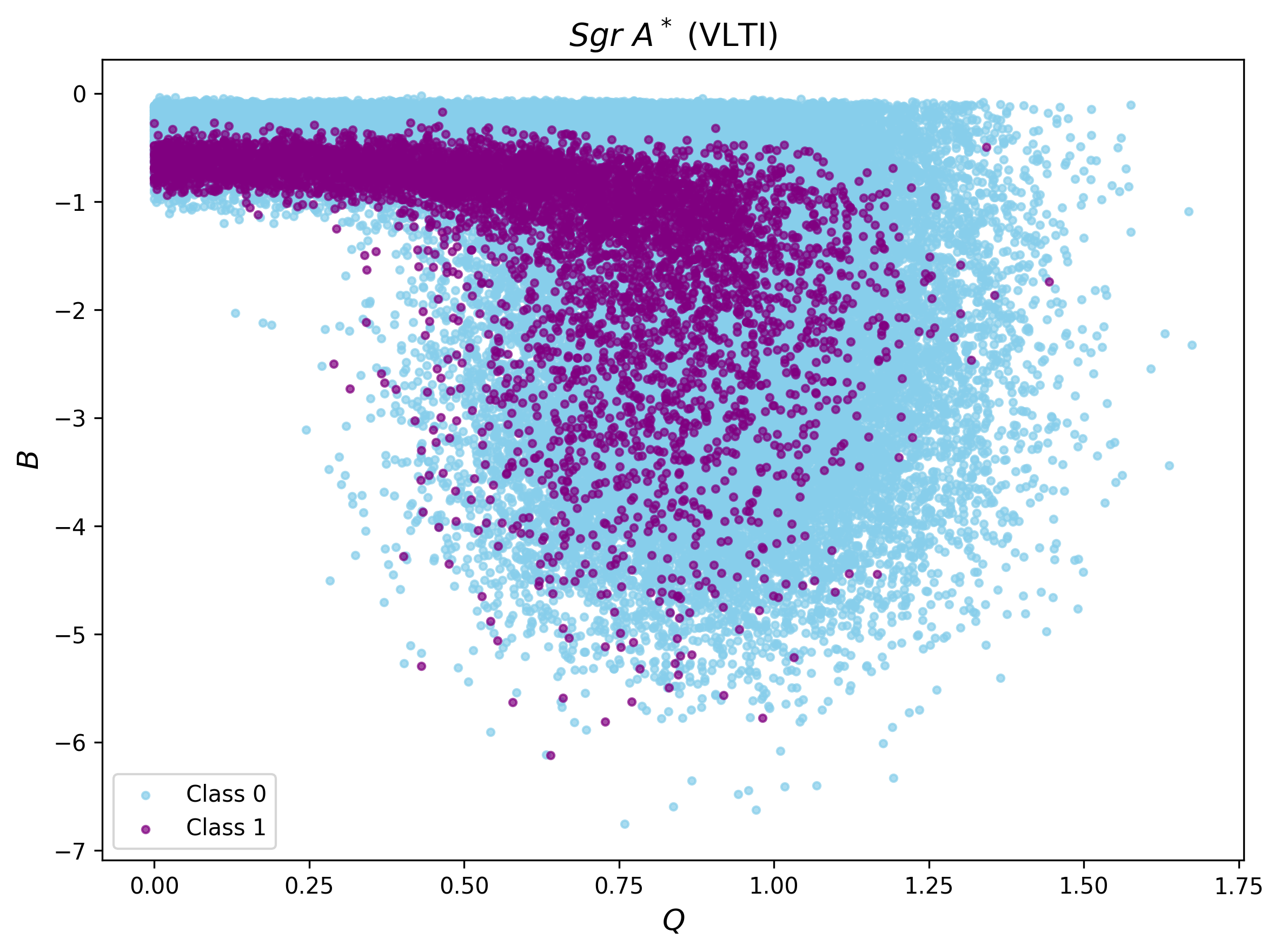}\hspace{2mm}
\includegraphics[scale=0.26]{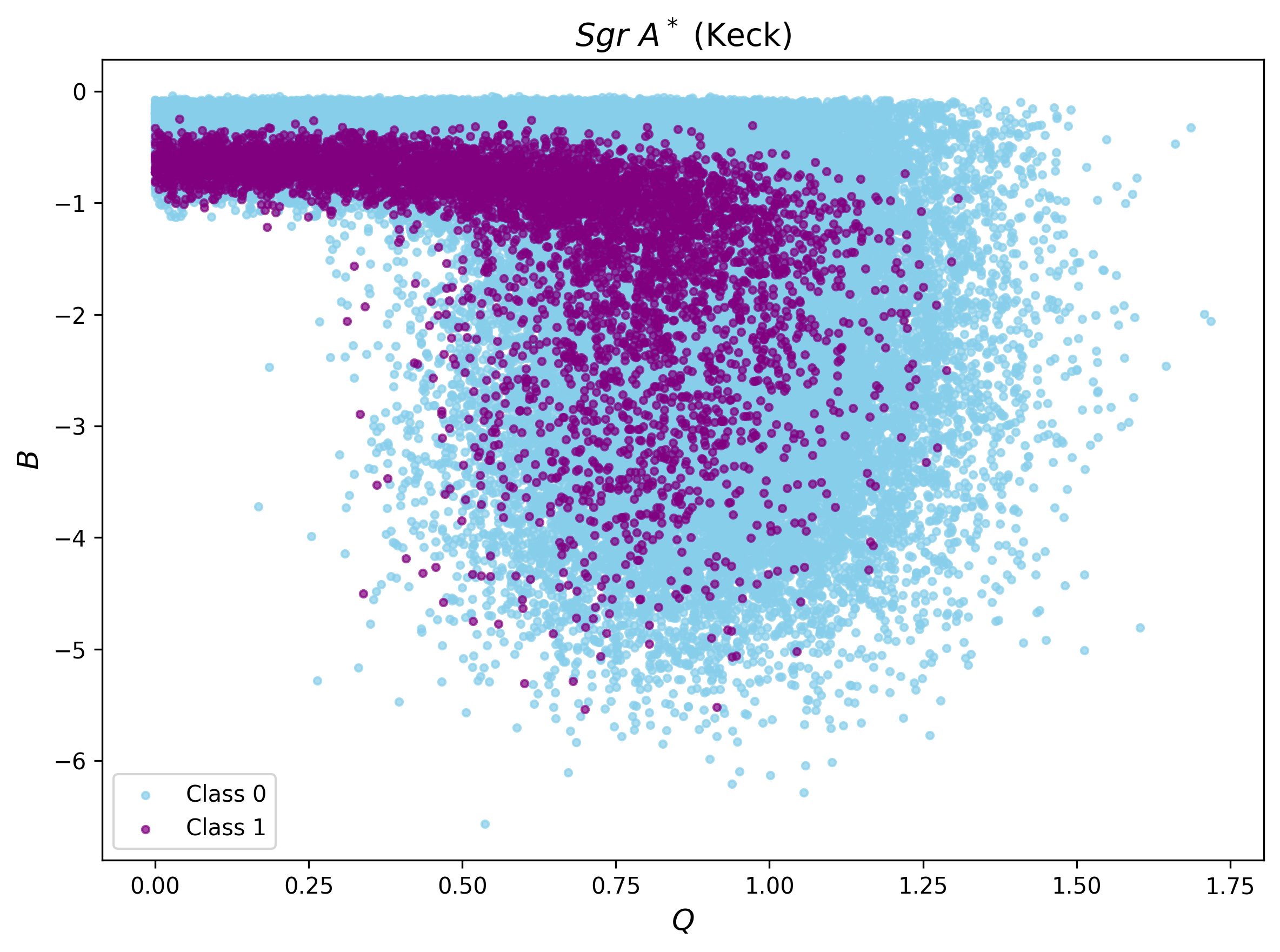}
\caption{Distribution of the compatible (Class 1) and incompatible (Class 0) black hole shadow configurations in the $(B,Q)$ parameter space for M87*, Sgr*A$_{\mathrm{VLTI}}$, and SgrA*$_{\mathrm{Keck}}$ at the $1-\sigma$ (top) and $2-\sigma$ (bottom) confidence levels.}
\label{BQ}
\end{figure}
The resulting distributions in the $(B,Q)$ parameter space are shown in Fig.~(\ref{BQ}), where the configurations are distinguished according to their binary classification. Class 1 corresponds to the  black hole shadow configurations compatible with the corresponding EHT observational constraints, while  the Class 0 denotes incompatible configurations. The two classes exhibit a significant overlap in the $(B,Q)$ parameter space, while their distributions show a clear dependence on the stringy parameter $B$ and the electric charge $Q$. The resulting structure varies with the considered observational dataset and confidence level, indicating a complex and nonlinear separation between compatible and incompatible configurations. This behavior provides a natural basis for formulating the problem as a binary classification task and motivates the subsequent application of the FCNN model.\\
 As in the thermodynamic study, non-physical and redundant configurations are removed in order to retain only physically significant parameter ranges. In fact,   the  resulting datasets are normalized  via  a standard scaling method to achieve better stability and convergence in the learning process. The data are then divided into training, validation, and test subsets. This  ensure   that each configuration appears in only one subset in order to  prevent data overfitting and guarantees a  reliable generalization. The FCNN architecture, dataset balancing procedure, and training strategy employed in this  part  are identical to those introduced in the previous thermodynamic analysis. The model is trained using the Adam optimizer together with a binary cross-entropy loss function. In particular, the neural network is trained using $5,832$ data samples. During the training phase, the learning rate is automatically adjusted to improve both the stability and the efficiency of the convergence process. The corresponding training behaviors are illustrated in Figs.~(\ref{fig:train}) and (\ref{fig:train1}) for the $1-\sigma$ and $2-\sigma$ observational constraints, respectively.

As shown in these figures, the loss decreases rapidly during the initial epochs before progressively stabilizing, indicating efficient learning and early convergence. Moreover, both the training and the  validation accuracies remain consistently high throughout the learning process, demonstrating a good generalization capability without noticeable overfitting. Small fluctuations observed in the validation loss are expected and originate from the stochastic mini-batch optimization procedure. However, these fluctuations do not affect the  associated  convergence behaviors. The trained model is further evaluated using independent test datasets. The accuracy is computed separately for each observational case.  The obtained results confirm the robustness of the classifier across all configurations. In particular, the network successfully distinguishes between configurations being  compatible and incompatible with the EHT observational constraints. This  shows  that it has learned a well-defined decision boundary in the parameter space  corresponding to   the  stringy parameter. The detailed performance metrics are summarized in Table~(\ref{s11}) for both the $1-\sigma$ and $2-\sigma$ observational constraints.

\begin{table}[!ht]
\centering
\caption{Model performance of the FCNN model.}
\label{tab11}

\begin{tabular}{lllll}
\toprule
{Metric} & {Confidence} & M87* & {SgrA*$_{\mathrm{Keck}}$} & {SgrA*$_{\mathrm{VLTI}}$} \\
\midrule

\multirow{2}{*}{Test Accuracy}
& $1-\sigma$ & 0.9892 & 0.9898 & 0.9903 \\
& $2-\sigma$ & 0.9926 & 0.9888 & 0.9898 \\

\bottomrule
\end{tabular} \label{s11}

\end{table}

To further evaluate the robustness of the classifier, we implement a voting strategy inspired by symmetry-based classification methods. For each configuration, 100 perturbed samples preserving the underlying symmetry are generated and classified independently. The final prediction is then obtained through  a majority voting. This procedure permits testing the mode stability in the presence of structured perturbations. The results show high and consistent voting accuracy across all datasets. This confirms that the classifier remains stable against small deformations that preserve the symmetry while effectively generalizing beyond the training distribution. Although global indicators such as test accuracy and voting accuracy provide a general assessment of the model's performance, they do not offer detailed information about the distribution of classification results.  To address this issue, we analyze the confusion matrices presented in Table~(\ref{c11}) for the $1-\sigma$ and $2-\sigma$ observational constraints, respectively. These matrices provide a more detailed description of how the model assigns predictions across the two classes by explicitly displaying correct classifications and misclassifications. In particular, they make it possible not only to quantify the number of prediction errors but also to identify potential imbalances between the classes. The results clearly show that the model correctly classifies nearly all samples, with only a very small number of misclassifications. Consequently, the confusion matrices provide a more transparent and comprehensive evaluation of the classifier performance beyond the aggregate accuracy metrics.

\begin{table}[!ht]
\centering
\caption{Confusion matrix for FCNN model  with  $1-\sigma$ and $2-\sigma$ observational constraints  at the  Van der Waals limits.}
\label{c11}
\begin{tabular}{l|l|c|c|c}
\toprule
\textbf{} & \textbf{Class} & \textbf{$M87^*$} 
& \textbf{$SgrA^*_{\mathrm{VLTI}}$} & \textbf{$SgrA^*_{\mathrm{Keck}}$} \\
\midrule
\multirow{2}{*}{$1-\sigma$} 
&  & Class 0 \quad Class 1 &  Class 0 \quad Class 1 &  Class 0 \quad Class 1 \\
& Class 0 &5,353 \qquad  38&5,540\qquad 12 &5,514 \qquad 38  \\
& Class 1 &25 \qquad 416& 47 \qquad 233 &18 \qquad 262 \\
\midrule
\multirow{2}{*}{$2-\sigma$} 
&  & Class 0 \quad Class 1 &  Class 0 \quad Class 1 &  Class 0 \quad Class 1 \\
& Class 0 &4,760 \qquad 18&5,245\qquad 17 &5,540 \qquad 12  \\
& Class 1 & 25 \qquad 1,029 &48 \qquad 522&47 \qquad 233  \\
\bottomrule
\end{tabular}
\end{table}

A detailed comparison reveals that the $2-\sigma$ observational constraint provides slightly more stable classification results than the $1-\sigma$ case, particularly for the M87* dataset, where the highest accuracies are obtained. In contrast, the $1-\sigma$ constraint introduces a broader uncertainty range, leading to a slight reduction in the classification performance. Nevertheless, the model remains highly robust under both observational regimes. These results indicate that the classifier is subject to relevant constraints under the $2-\sigma$ observation framework, while maintaining excellent generalization capabilities in the more lenient $1-\sigma$ scenario.

\section{Conclusions and  concluding remarks}
Using machine learning methods, we have investigated both the thermodynamic and the  optical properties of  $B$-deformed RN-AdS  black holes inspired by string theory and related frameworks, including NC  geometry. We have first analyzed their thermodynamic behaviors  by exploring the Van der Waals criticality and the associated universal relations in terms of the  stringy parameter $B$. With the help of  advenced numerical computations, we have established an  interplay between the black hole parameters governing   the $P$--$V$ criticality and the  Joule–Thomson expansion features.  The generated datasets have subsequently been implemented within a FCNN, which has accurately identified black hole configurations exhibiting Van der Waals phase transitions. The obtained  results, of the present work, have demonstrated    potential applications  of machine learning methods. It has been considered  as an efficient tool for constraining black hole parameters  via certain  thermodynamic observables.

After that, we  have then  studied the optical properties of these black holes by analyzing their shadows using such numerical methods.   Performing  appropriate computations,   we have observed that the electric charge $Q$, the cosmological constant $\Lambda$, and the stringy  parameter $B$ have affected only the size of the black hole shadow geometries, while its circular shape has remained unchanged since we are considering non-rotating solutions. More precisely, we have found that the  $\Lambda$ and $B$ parameters  have exhibited similar effects on the shadow radius, whereas the electric charge $Q$ has shown  an  opposite behavior.  Exploiting  the  the shadow observational data  reported by the EHT  international collaborations, we have constrained the stringy  parameter  $B$. In particular, we have found that  such a parameter  has remained negative and below a certain upper bound to be consistent with the observational data. The shadow datasets  being generated from such  numerical simulations have also been implemented within a FCNN framework to establish a possible  connection between the  obtained theoretical predictions and the shadow observations reported by the EHT international  collaborations.  It has been remarked that the trained FCNN has accurately identified the observationally allowed shadow configurations for both M87* and Sgr A* black hole  models. A comparative analysis has revealed that the $2-\sigma$ observational constraint has provided slightly better and more stable classification performance than the $1-\sigma$ case, particularly for the M87* dataset. Under the $1-\sigma$ constraint, we have remarked  that the larger uncertainty range has slightly reduced the classification performance due to  the increased overlap between the allowed and  the excluded parameter regions.

This work  comes up with  some  possible  directions   for future investigations. Recently, the black hole techniques  have emerged in the field of condensed matter physics   via  similarities and analogous scenarios  to approach semi-metals \cite{DFT1,DFT2}. This subject  may open application gates for machine learning and advanced numerical methods  in the density functional theory (DFT) where the thermodynamic  and the optical properties of materials  have been extensively studied using different numerical methods  including  the quantum espresso code   with certain approximations\cite{anas}.  In this way, it  could be possible  to  establish a nice interplay between  black hole physics and  condensed matter physics  by implementing  DFT methods  in  gravity equations. This link  between  semimetals and black holes could  be exploited to develop  complex issues  including  the Hawking radiation and the quantum information scenarios. Alternatively,  a  natural extension of the present work   could be  to develop a complete  machine learning framework to obtain stronger constraints on the  black hole parameters in generalized  models  of gravity theory. It would also be interesting  to establish a connection with recent machine learning approaches devoted to constraint the cosmological constant  via  thermodynamic properties, as discussed in ~\cite{35}. We expect that these investigation  directions will provide further insights into the bridging scenarios  between black hole physics, gravitational theories, and  more advanced numerical methods  including parallel programing  computations.\\

{\bf Data availability}\\
  The data are available from the corresponding author upon reasonable request. \\
  
{\bf Funding information}\\
  It is  not applicable.


\section*{Acknowledgements}
 MJ would like to thank S. E. Baddis,    H. Belmahi  and  S.E. Ennadifi   for  collaborations  on related topics. Her   work has been supported  by CNRST in the frame of the PhD Associate Scholarship Program PASS.

\end{document}